\documentclass[pdftex,a4paper,sort&compress]{scrartcl}
\usepackage[english]{babel}
\usepackage[utf8]{inputenc}
\usepackage[T1]{fontenc}

\usepackage{tabularx}

\usepackage{multicol}
\usepackage{csquotes}
\usepackage{amsmath,amssymb}
\usepackage{amsthm}
\usepackage{thmtools} % https://tex.stackexchange.com/a/732209
\usepackage[pdfusetitle]{hyperref}
\usepackage{mathtools}
\usepackage{upgreek}
\usepackage{nicefrac}
\usepackage{slashed}
\usepackage{physics}
\usepackage{xcolor}
\usepackage{faktor}
\usepackage{shuffle}
\usepackage{siunitx}
\usepackage{booktabs}
\usepackage{upquote}
\usepackage[nameinlink,noabbrev]{cleveref}

\numberwithin{equation}{section}
\usepackage[sort&compress,numbers]{natbib}
\theoremstyle{definition}
\newcommand{\inv}[1]{\frac{1}{#1}}
\newcommand{\C}{\mathbb{C}}

\newcommand{\Z}{\mathbb{Z}}

\newcommand{\fullstop}{\text{\,.}}
\newcommand{\comma}{\text{\,,}}
\DeclareMathOperator{\res}{Res}

\def\beq{\begin{equation}}
\def\eeq{\end{equation}}
\def\bsp#1\esp{\begin{split}#1\end{split}}

\DeclareMathOperator{\dlog}{dlog}

\newcommand{\IterInt}{\textsc{IterInt}}
\newcommand{\ginac}{\textsc{GiNaC}}
\newcommand{\mathematica}{\textsc{Mathematica}}
\newcommand{\boost}{\textsc{Boost}}
\newcommand{\GSL}{\textsc{GSL}}
\def\cpp{{C\nolinebreak[4]\hspace{-.05em}\raisebox{.4ex}{\tiny\textbf{++}}}}

\begin{document}

\thispagestyle{empty}

\begin{flushright}
BONN-TH-2026-13
\end{flushright}

\vspace{1.5cm}

\begin{center}
  {\Large\textbf{\textsc{IterInt}: Evaluating iterated integrals via differential equations}\\
  }
  \vspace{1cm}
  {\large Gideon Baur$^{a}$, Claude Duhr${}^{a,b}$} \\
  \vspace{1cm}
      {\small \em ${}^{a}$ Bethe Center for Theoretical Physics, Universit\"at Bonn, D-53115 Bonn, Germany}\\
       \vspace{2mm}
      {\small \em ${}^{b}$ Cluster of Excellence ``Color meets Flavor'', Universit\"at Bonn, D-53115 Bonn, Germany.} \
\end{center}

\vspace{2cm}

% abstract ---------------------------------------
\begin{abstract}\noindent
  {
\textsc{Abstract:} We introduce \IterInt, a novel package implemented in both \mathematica\ and \cpp\ for the numerical evaluation of iterated integrals involving arbitrary integration kernels. After the user has defined the integration kernels, \IterInt\ transforms the iterated integrals into a system of first-order linear differential equations which can be solved efficiently and with high precision using well established libraries. \IterInt\ is also able to automatically perform shuffle-regularisation. This makes it possible to evaluate also integrals where the integrand has a pole at the starting point of the integration path. As an illustration of our code, and also to validate it and gauge its performance, we compare the output of \IterInt\ to the results obtained by \ginac\ for ordinary and elliptic multiple polylogarithms, and also to existing results for the first few orders for banana integrals with up to four loops. 
   }
\end{abstract}

\vspace*{\fill}

% -----------------------------------------------------------------------------

\newpage

\subsection*{Program summary}
\begin{small}

{\em Program Title:}  \IterInt\    
                               \\

\noindent{\em CPC Library link to program files:} (to be added by Technical Editor) \\

\noindent{\em Developer's repository link:} \url{https://github.com/baugid/IterInt} \\

\noindent{\em Licensing provisions(please choose one):} GNU Public License.\\

\noindent{\em Programming language:} \cpp\ and \mathematica\                                 \\

\noindent{\em Supplementary material:}\\

\noindent{\em Nature of problem:} Many function classes, including multiple polylogarithms, appearing in the study of Feynman integrals can be expressed as Chen iterated integrals.
  These functions are defined by repeated integration along a path. Iterated integrals may quickly become difficult to numerically evaluate with textbook methods due to the high dimension of the integration domain.
  While there are publicly available algorithms that tackle this in certain special cases, there is no public algorithm solving the general problem.\\
  
\noindent{\em Solution method:}
By construction (iterated) integrals can be equivalently expressed as the solution of a system of ordinary differential equations.
In this way a $N$-dimensional integration is converted into a system of $N+1$ coupled differential equations. Such a system can then be tackled using standard methods.
As the behaviour of such algorithms is generally much better as $N$ increases, the usual \emph{curse of dimensionality} is avoided.
\\

\noindent{\em Additional comments including restrictions and unusual features:} When studying Feynman integrals, the resulting iterated integrals generally diverge and therefore require regularisation.
Our code implements an algorithmic way to perform this regularisation.
In general our algorithms adapt to a wide array of integrands without external intervention, especially including functions without globally-converging series expansions.
   \\
\end{small}

\newpage

\tableofcontents

% !TEX root = main.tex

\section{Introduction}
\label{sec:intro}

Iterated integrals, defined as the integration of a sequence of a one-forms along a path in a given order, play an extremely important role both in pure mathematics and in physics. In mathematics, following the seminal work by Chen~\cite{Chen:1977oja}, they describe the periods of the pro-unipotent completion of the fundamental group of a space. In physics, they arise when solving differential equations satisfied by dimensionally-regulated~\cite{tHooft:1972tcz,Bollini:1972ui} Feynman integrals~\cite{Kotikov:1990kg,Kotikov:1991hm,Kotikov:1991pm,Gehrmann:1999as,Henn:2013pwa}, and as such they are the cornerstone of almost all precision computations in collider physics and gravitational wave physics (see, e.g., ref.~\cite{Bourjaily:2022bwx} for a recent review). Hence, having a good understanding of iterated integrals, including efficient algorithms for their numerical evaluation, is paramount if we want to meet the precision requirements of current and future experiments.

In some instances, iterated integrals can be expressed in terms of known special functions. The arguably most prominent example of this are iterated integrals of dlog-forms with rational functions as arguments, which can always be expressed in terms of multiple polylogarithms~\cite{Goncharov:1998kja} (see also refs.~\cite{Gehrmann:2000zt,Remiddi:1999ew}). This class of special functions is well understood, cf.,~e.g.,~refs.~\cite{Goncharov:2010jf,Brown:2009qja,Duhr:2011zq,Duhr:2012fh,Duhr:2014woa,Weinzierl:2022eaz}, and there are several public libraries that allow one to evaluate (certain classes of) them~\cite{Gehrmann:2001pz,Gehrmann:2001jv,Vollinga:2004sn,Buehler:2011ev,Frellesvig:2016lxm,Naterop:2019xaf}. It is possible to extend polylogarithms to elliptic curves~\cite{Brown:2011wfj,Broedel:2014vla,Broedel:2017kkb,Enriquez2023EllipticH}. Also these functions are starting to be well understood, with first public packages to evaluate them having become available~\cite{Walden:2020odh,Duhr:2026ell}.

Ordinary and elliptic polylogarithms, however, do not yet exhaust the class of iterated integrals required to express Feynman integrals starting from two loops (not even if we focus on iterated integrals of dlog forms, cf., e.g., ref.~\cite{Duhr:2020gdd}). For example, it is known that certain Feynman integrals can be expressed in terms of iterated integrals of (meromorphic) modular forms~\cite{ManinModular,Brown:2014pnb,Matthes2022,Broedel:2021zij}. More generally, also iterated integrals over integration kernels involving the (quasi-)periods of Calabi-Yau varieties appear (cf.,~e.g., refs.~\cite{Bonisch:2021yfw,Pogel:2022ken,Pogel:2022vat,Pogel:2022yat,Duhr:2025lbz,Klemm:2024wtd,Driesse:2024feo,Dlapa:2025biy}). There are also Feynman integrals associated to higher-genus Riemann surfaces~\cite{Huang:2013kh,Hauenstein:2014mda,Marzucca:2023gto,Duhr:2024uid,Bargiela:2025vwl,Yang:2026rgb}, and it is expected that these Feynman integrals may be evaluated using extensions of multiple polylogarithms to higher-genus curves~\cite{DHoker:2023vax,DHoker:2024ozn,Baune:2024biq,Baune:2024ber,DHoker:2025szl,DHoker:2025dhv,Baune:2025sfy,DHoker:2026lgg,DHoker:2026ggx,Berger:2026bwv}. For such classes of special functions and iterated integrals, there are currently no public packages for their numerical evaluation (though in some cases, in particular when modular forms are involved, public packages exist~\cite{Walden:2020odh,Prausa:2020psw}). More generally, modern approaches to solving differential equations for Feynman integrals typically lead to iterated integrals involving differential forms that cannot immediately be expressed in terms of classes of special functions that have already been studied in the literature~\cite{Pogel:2022ken,Pogel:2022vat,Pogel:2022yat,Gorges:2023zgv,Duhr:2025lbz,Forner:2026vby}.
The numerical routines for these functions then need to be painstakingly implemented on a case by case basis. Having a flexible framework to evaluate very general classes of iterated integrals with high precision, without specific requirements on the integration kernels or the classes of special functions to which they may evaluate, is therefore highly desirable, but it is currently still lacking.

In this paper, we introduce the package \IterInt, which allows the user to numerically evaluate very general classes of iterated integrals. The user only needs to implement numerical routines to evaluate the integration kernels, which is typically a much simpler task. \IterInt\ then provides the algorithms to automatically evaluate iterated integrals over these kernels. To this effect, \IterInt\ transforms the iterated integrals into a system of first-order linear differential equations, which can be solved numerically in an efficient way using well-established numerical libraries. Since in applications (to Feynman integrals) one is typically not interested in evaluating just a single iterated integral, but possibly hundreds, or even thousands, of them, \IterInt\ optimises the system of differential equations such as to avoid a redundancy in the iterated integrals that need to be evaluated. Moreover, the starting point of the iterated integration is often a singular point. In that scenario the iterated integrals needs to be interpreted as suitably shuffle-regularised versions. \IterInt\ implements the combinatorial formulation of shuffle-regularisation from ref.~\cite{Brown:2014pnb}, which allows it to easily perform the regularisation and apply its numerical algorithms to convergent integrals.

We have implemented the algorithms of \IterInt\ in both \mathematica\ and \cpp\ (and for \cpp\ we provide two different backends, based on the \boost\ and \GSL\ libraries, respectively). As a validation of our code, and as  a means to gauge its performance and precision, we have compared the results of \IterInt\ to the existing implementations of ordinary and elliptic multiple polylogarithms into \ginac~\cite{Bauer:2000cp}. We have also implemented the iterated integrals required to evaluate banana integrals with up to four loops (for specific mass configurations) and compared them to results in the literature.

Our paper is organised as follows: In section~\ref{sec:iterated} we provide a brief review of iterated integrals, focusing in particular on shuffle-regularisation and homotopy-invariance. In section~\ref{sec:algo} we present our main result, namely we describe our algorithm to evaluate iterated integrals via differential equations, and we discuss various optimisations one can perform. In section~\ref{sec:codes} we describe the usage of the \mathematica\ and \cpp\ implementations of \IterInt. Finally, in sections~\ref{sec:examples_MPLs} and~\ref{sec:examples} we validate our code and discuss its performance for ordinary and elliptic polylogarithms and for banana integrals with up to four loops. In section~\ref{sec:conc} we draw our conclusions. We also include an appendix where we show how to relate two different regularisation schemes for elliptic polylogarithms.

% !TEX root = main.tex

\section{Review of iterated integrals}\label{sec:iterated}
Before we present our algorithm and its implementation in subsequent sections, we review some background on iterated integrals. 

\subsection{Definition and basic properties}

Consider a space $X$, which for us is typically some subset of the $N$-dimensional complex space $\mathbb{C}^N$, and a path $\gamma: [0,t] \to X$. We also consider differential one-forms $\omega_i$ on $X$. We can pull these one-forms back to the path $\gamma$, and we write $\gamma^*\omega_i=f_i(\xi) \dd{\xi}$, where $\xi\in[0,t]$ is a coordinate on $\gamma$ and $f_i(\xi)$ are complex-valued functions.
The iterated integral of a sequence of differential one-forms along the path $\gamma$ is then defined as~\cite{Chen:1977oja}:
\begin{align}
I_{\gamma}(\omega_1,\ldots,\omega_n) =  \int_\gamma \omega_1\cdots\omega_n = \int_{0\leq \xi_1\leq\cdots\leq \xi_n\leq t}\dd{\xi_n}\dots\dd{\xi_1} f_n(\xi_n)\cdots f_1(\xi_1)\,.\label{eq:defIter}
\end{align}
The sequence $\omega_1\cdots\omega_n$ is often referred to as a \emph{word}, and the individual differential forms $\omega_i$ are called the \emph{letters}. The number $n$ of letters in a word is called its \emph{length}. By convention, the iterated integral over the empty word (of length zero) is defined as 1.

Iterated integrals enjoy various properties~\cite{Chen:1977oja}. The most prominent property is the shuffle product, which allows one to write a product of iterated integrals over the same path $\gamma$ as a linear combination of iterated integrals:
\begin{align}\label{eq:shuffle-product}
    I_{\gamma}(\omega_1,\dots,\omega_k)\, I_{\gamma}(\omega_{k+1},\dots,\omega_n) & =\sum_{\sigma\in S(k,n)}I_{\gamma}(\omega_{\sigma(1)},\dots,\omega_{\sigma(n)})\,,
\end{align}
where the sum runs over the set $S(k,n)$ of all shuffles of two words with $k$ and $n-k$ letters, i.e., the set of all permutations of $n$ elements that leave the order of the letters in each of the two words separately intact. Iterated integrals also behave nicely under composition and reversal of paths,
\beq\bsp\label{eq:path-composition}
I_{\gamma_1\gamma_2}(\omega_1,\ldots,\omega_n) &\,= \sum_{k=0}^nI_{\gamma_1}(\omega_1,\ldots,\omega_k) \,I_{\gamma_2}(\omega_{k+1},\ldots,\omega_n) \,,\\
I_{\gamma^{-1}}(\omega_1,\ldots,\omega_n) &\,=(-1)^n\,I_{\gamma}(\omega_n,\ldots,\omega_1)\,,
\esp\eeq 
where $\gamma^{-1}$ is the path $\gamma$ traversed in the opposite direction and $\gamma_1\gamma_2$ is obtained by first traversing $\gamma_1$ and then $\gamma_2$, where $\gamma_1$ and $\gamma_2$ are two paths such that the end-point of $\gamma_1$ agrees with the starting point of $\gamma_2$.

It can be useful to understand these three properties of iterated integrals as operations on words only. To this effect, consider the vector space $\mathcal{B}$ (say, over the complex numbers) generated by all words built from a (finite) set of letters $\omega_i$. This vector space can be turned into Hopf algebra, where the multiplication encodes the shuffle product in eq.~\eqref{eq:shuffle-product} and the coproduct and the antipode encode the path composition and reversal formulas from eq.~\eqref{eq:path-composition}. The multiplication on $\mathcal{B}$ is simply the shuffle product of words,
\begin{align}
    \qty(\omega_1\cdots\omega_k)\shuffle \qty(\omega_{k+1}\cdots\omega_n) & =\sum_{\sigma \in S(k,n)}\omega_{\sigma(1)}\cdots\omega_{\sigma(n)}\,.
\end{align}
The coproduct is given by the deconcatenation of words
\beq\bsp
\Delta(\omega_1\cdots\omega_n) &\,= \sum_{k=0}^n\omega_1\cdots\omega_k\otimes\omega_{k+1}\cdots\omega_n \,,
\esp\eeq 
and the counit $\epsilon:\mathcal{B}\to\mathbb{C}$ is the projection onto the empty word. The antipode is given by the reversal of words, up to a sign,
\beq\bsp
S(\omega_1\cdots\omega_n) &\,=(-1)^n\,\omega_n\cdots\omega_1\,.
\esp\eeq 
It is easy to check that these operations satisfy all the conditions to turn $\mathcal{B}$ into a Hopf algebra. If $f,g:\mathcal{B}\to \mathbb{C}$ is an algebra homomorphism from $\mathcal{B}$ to $\mathbb{C}$,\footnote{We could work over any ring $R$ and consider characters with values in $R$.} we can define their convolution product
\beq
f\ast g = m(f\otimes g)\Delta\,,
\eeq
where $m$ is the usual multiplication in $\mathbb{C}$. It can be shown that the Hopf algebra structure on $\mathcal{B}$ induces a group structure on algebra-homomorphisms whose group law is the convolution product. In particular, the unit for the convolution product is the counit $\epsilon$ on $\mathcal{B}$ and the inverse is composition with the anitpode, $f^{\ast-1}=fS$.

\subsection{Homotopy-invariance}

A priori the iterated integrals in eq.~\eqref{eq:defIter} are functions of the details of the path $\gamma$. In applications, one is typically interested in \emph{homotopy-invariant} iterated integrals, which are functions of the end-points of the path only. More precisely, homotopy-invariant iterated integrals are functions of  the homotopy class of $\gamma$ in $X$, and the integral is invariant under deformations of the path that keep its end-points fixed and where the deformation does not cross any singularities (poles or branch cuts) of the letters.

Even though individual letters may not lead to iterated integrals that are homotopy-invariant, it is possible to describe linear combinations that are. Such linear combinations are called \emph{integrable}. For iterated integrals of length one (i.e., ordinary integrals), homotopy-invariance reduces to the condition that the differential form is closed, $\dd\omega_1=0$. For higher length, this condition can be generalised. A linear combination $\rho$ of words is integrable if and only if it satisfies $D\rho=0$, where the differential acts on words via the formula~\cite{Chen:1977oja}
\beq\label{eq:integrability}
D(\omega_1\cdots \omega_n) = \sum_{k=1}^n \omega_1\cdots (\dd\omega_k)\cdots \omega_n + \sum_{k=1}^{n-1} \omega_1\cdots (\omega_{k}\wedge\omega_{k+1})\cdots\omega_n \,.
\eeq 

Iterated integrals naturally arise from the computation of multiloop Feynman integrals from differential equations (see section~\ref{sec:examples}).
Note that iterated integrals that arise as solutions to an integrable system of first-order differential equations (i.e., a system defined by a flat connection), the resulting linear combinations are always integrable.
Moreover, in many applications the differential forms $\omega_k$ are closed. In that case the integrability condition in eq.~\eqref{eq:integrability} simplifies, and only the second term contributes. We recover in this way the well-known integrability condition~\cite{Gaiotto:2011dt} of symbols for dlog-forms~\cite{Chen:1977oja,Goncharov:2010jf,Brown:2009qja,Duhr:2011zq}.

In the following we always assume that our (combinations of) iterated integrals are homotopy-invariant, and we interpret them as functions of the end-point $\gamma(t)$ of the path $\gamma$. To this effect, we introduce the notation 
\beq
I(\omega_1,\dots,\omega_n; t) =  I_{\gamma}(\omega_1,\dots,\omega_n) = \int_{0}^{t}\omega_1\cdots\omega_n\,.
\eeq
For applications and after pulling back to the path $\gamma$, it is typically sufficient to have tools to evaluate the integrals seen as a function of the single complex variable $t$. We will therefore from now on exclusively focus on iterated integrals defined on a space $X$ of complex dimension $N=1$. 
Note that in this setup the homotopy-invariance is automatic, if for $N=1$ the wedge product of any two holomorphic one-forms vanishes. Hence, since we focus on $N=1$, all iterated integrals are individually homotopy-invariant, and we may assume without loss of generality that $\gamma$ is the straight line segment in the complex plane from $0$ to $t$.

\subsection{Regularisation}
\label{sec:regularisation}
The iterated integral in eq.~\eqref{eq:defIter} is convergent whenever the path $\gamma$ does not pass through any singularity of the letters $\omega_i$. If some $\omega_i$ has a pole\footnote{The forms may have integrable singularities. These may lead to numerical instabilities, but the integrals are formally convergent.} on the interior of $\gamma$, i.e., at some point $\gamma(\xi)$ with $\xi\in (0,t)$, then we may slightly deform the contour away from the singularity. The value of the integral then depends on how precisely we deform the contour. In the following we assume that the contour was chosen in such a way that there are no poles on the interior of $\gamma$.

If some of the letters have poles at one of the end points, the contour cannot be deformed, and the integral in eq.~\eqref{eq:defIter} will typically diverge. In particular, if $f_1(\xi)$ has a pole at the initial point $\xi=0$, then the iterated integral diverges for all values of $t$. In applications, the point $\xi=0$ is usually related to the initial condition of the differential equation satisfied by dimensionally-regulated Feynman integrals, which is often easiest to obtain at a singular point of the differential equation. It is therefore important to extend the definition in eq.~\eqref{eq:defIter} to include versions of iterated integrals that are suitably regularised at the initial point $\xi=0$.

If we focus on iterated integrals that arise from canonical differential equations for Feynman integrals, we expect that the differential forms $\omega_i$ have at most simple poles at $\xi=0$. 
In that case a convenient regularisation is the so-called \emph{shuffle} or \emph{tangential base-point regularisation}, cf.,~e.g.,~ref.~\cite{Deligne1989}.\footnote{For an extension to higher-order poles via renormalisation, see ref.~\cite{Matthes2022}.} In a nutshell, it can operationally be described as follows. We shift the lower integration boundary to $\xi=\varepsilon\neq0$, and we consider the resulting integral as a function of $\varepsilon$,
\begin{align}\label{eq:hatIDefinition}
\hat{I}(\varepsilon,t) :=   \int_{\varepsilon}^{t}\dd{\xi_n}\,f_n(\xi_n)\,\int_{\varepsilon}^{\xi_n}\cdots  \int_{\varepsilon}^{\xi_2}\dd{\xi_1}\,f_1(\xi_1)\,.
\end{align}
If $f_1(\xi)$ is regular at $\xi=0$, then $\hat{I}(\varepsilon,t)$ defines an analytic function of $\varepsilon$ at $\varepsilon=0$, and the limit $\varepsilon\to0$ is smooth. If, however, $f_1(\xi)$ has single pole at $\xi=0$, then the integral diverges logarithmically. In a small neighborhood around $\varepsilon=0$, $\hat{I}(\varepsilon,t)$ may be cast in the form
\beq
\hat{I}(\varepsilon,t)= \sum_{r=0}^p\hat{I}_r(\varepsilon,t)\,\log^r\varepsilon\,,
\eeq
where the $\hat{I}_r(\varepsilon,t)$ are analytic at $\varepsilon=0$, i.e., they admit a Taylor series expansion around $\varepsilon=0$. The (shuffle-)regulated value of $I(\omega_1,\ldots,\omega_n;t)$ is then defined by first putting all $\log\varepsilon$ in $\hat{I}(\varepsilon,t)$ to zero, and then taking the limit $\varepsilon\to0$ (cf.,~e.g.,~ref.~\cite{Brown:2008um}):
\beq\label{eq:regulator_def_eps}
\operatorname{Reg}_v[\hat{I}(\varepsilon,t)]:=\hat{I}_0(0,t)\,.
\eeq
Here $v$ is a non-zero complex number, which can be understood as a choice of regularisation scheme defined by rescaling $\varepsilon$ to $v\varepsilon$. Its impact on the regularisation will become clear below.

While eq.~\eqref{eq:regulator_def_eps} provides a rigorous definition of the regularised versions of the iterated integrals, it is hard to implement eq.~\eqref{eq:regulator_def_eps} in practice, because one needs to evaluate the integrals for generic values of $\varepsilon$ (in a neighborhood of the origin) and then take the limit. In particular, such a procedure is extremely hard to implement into a numerical code. We therefore rely  on another, equivalent, formulation of shuffle regularisation introduced in ref.~\cite{Brown:2014pnb}. We first extract for every one-form $\omega_i$  its behaviour close to the origin,
\beq
a_i:= \res_{\xi=0}\dd \xi\,f_i(\xi)\textrm{~~~and~~~}\omega_i^\infty:=a_i\dlog \xi\,.
\eeq
It is then possible to show that the shuffle-regularisation can be encoded into purely algebraic manipulations on words: 
\begin{align}
    \operatorname{Reg}_v\int_\gamma \omega_1\cdots \omega_n & =\sum_{k=0}^{n}\inv{k!}\log^k\frac{t}{v}\,\qty(\prod_{j=1}^k a_j) \int_\gamma R[\omega_{k+1}\dots\omega_n]\comma\label{eq:regularize}
\end{align}
where $R$ denotes a purely combinatorial operation on words formed from letters $\omega_i$,
\begin{align}
    R[\omega_1\dots\omega_l] & :=\sum_{k=0}^{l}{\qty(-1)}^k \qty{\qty(\omega_k^\infty\dots\omega_1^\infty)\shuffle\qty(\omega_{k+1}\dots\omega_l)}\fullstop
\end{align}
This map can be succinctly written as a convolution product in the Hopf algebra of words,
\beq\label{eq:regularisaiton_Hopf}
R = \pi_{\infty}^{\ast-1}\ast \operatorname{id} = \pi_{\infty}S\ast\operatorname{id}\,,
\eeq
where $\pi_{\infty}(\omega_i) = \omega_i^{\infty}$ extracts the singular behavior of the letter $\omega_i$.  Using the group structure of the convolution product, eq.~\eqref{eq:regularisaiton_Hopf} can be cast in the equivalent form ${\operatorname{id} = \pi_{\infty}\ast R}$. 

It can be shown that the integrals involving $R$ are absolutely convergent~\cite{Brown:2014pnb}, making them suitable for numerical integration. However, it is only the linear combination in eq.~\eqref{eq:regularize} that is integrable, and individual terms may still lead to divergent integrals. It is convenient to cast eq.~\eqref{eq:regularize} into a form where all terms are explicitly convergent. To this effect, we define the pole-free part of $\omega_i$ as 
\begin{equation}
\overline{\omega}_i=\omega_i-\omega_i^\infty\,.
\end{equation}
With this notation the expression $R[\omega_1\dots\omega_l]$ can be rewritten as
\begin{align}\label{eq:RCalculable}
    R[\omega_1\dots\omega_l]&=\sum_{k=0}^{l-1} {\qty(-1)}^k \qty{\overline{\omega}_{k+1}\qty[\qty(\omega_k^\infty\dots\omega_1^\infty)\shuffle\qty(\omega_{k+2}\dots\omega_l)]}\fullstop
\end{align}
As the first differential form $\overline{\omega}_{k+1}$ in each word does not have a pole at $\xi=0$, the resulting iterated integrals are individually convergent.
% !TEX root = main.tex

\section{Numerical evaluation via differential equations}\label{sec:algo}
In this section we present our algorithm to evaluate iterated integrals built from letters with at most simple poles, and we assume that there is no pole on the path of integration $\gamma$ (though we allow for poles at the starting point, which will be shuffle-regulated). We assume that the space $X$ on which the letters live is one-dimensional. This is not a restriction, because after fixing the path in a higher-dimensional space, we can effectively reduce the problem to one dimension after pulling everything back to the path. Hence, without loss of generality, we assume that all iterated integrals are evaluated over a straight line segment from 0 to a point $t$ in the complex plane. We also assume that $t$ is not a pole of $\omega_1$, because otherwise the integral fails to converge. From the discussion in the previous section, we know how to replace such integrals by a linear combination of convergent integrals, cf.~eq.~\eqref{eq:regularize}.

We now describe our strategy to evaluate such iterated integrals numerically. We start by focusing on a single absolutely convergent integral $I(\omega_1,\ldots,\omega_n;t)$, which implies in particular that $\omega_1$ has no pole at $t=0$. We will comment on the effect of the regularisation later in this section. $I(\omega_1,\ldots,\omega_n;t)$ then defines a differentiable function of the complex variable $t$ (as long as $t$ stays away from singularities of the $\omega_i$, including those at the end-point of the path).
The derivative of the iterated integral can directly be computed from eq.~\eqref{eq:defIter},
\begin{align}
    \dv{t} I(\omega_1,\dots,\omega_n; t) & = f_n(t)\,I(\omega_1,\dots,\omega_{n-1}; t)\fullstop
\end{align}
The right-hand side involves the known function $f_n$ and the yet unknown iterated integral $I(\omega_1,\dots,\omega_{n-1}; t)$. We may iterate this procedure to obtain a linear system of first-order differential equations,
\beq\label{eq:diffEq}
 \dv{t} I(\omega_1,\dots,\omega_k; t) = f_{k}(t)\,I(\omega_1,\dots,\omega_{k-1}; t)\,, \qquad0\le k\le n\,,
 \eeq
 where we defined $f_0(t)=0$. Equivalently, this system may be cast in a matrix form,
 \beq\label{eq:system_main_matrix}
 \dv{t}\mathbf{I}(t) = \mathbf{F}(t)\,\mathbf{I}(t) \,,
 \eeq
 where we defined
 \beq\bsp
 \mathbf{I}(t) &\,= \big(I(\omega_1,\ldots,\omega_{n-i};t)\big)_{0\le i\le n}\,,\\
 \mathbf{F}(t) &\,= \big(\delta_{j,i+1}\,f_{n+1-i}(t)\big)_{0\le i,j\le n}\,.
\esp \eeq
Note that, since the convergence of an iterated integral is controlled by the left-most letter $\omega_1$, we see that if $I(\omega_1,\ldots,\omega_n;t)$ is convergent  (and hence does not require regularisation), then the same is true for all other entries of the vector $\mathbf{I}(t)$. The initial condition of the system in eq.~\eqref{eq:system_main_matrix} comes from the fact that, if an iterated integral of length at least 1 is convergent, then it must vanish at $t=0$,
\beq
\lim_{t\to0}I(\omega_1,\ldots,\omega_k;t)=\left\{\begin{array}{ll}0\,, & \textrm{ if } k>0\,,\\
1\,,&\textrm{ if } k=0\,,\end{array}\right.
\eeq
or equivalently in vector form,
\beq\label{eq:init_cond}
\lim_{t\to0}\mathbf{I}(t) = \left(\begin{smallmatrix} 0\\\vdots\\0\\1\end{smallmatrix}\right)\,.
\eeq

We see that we can obtain the value of $I(\omega_1,\ldots,\omega_n;t)$ by solving the system in eq.~\eqref{eq:system_main_matrix} with the initial condition in eq.~\eqref{eq:init_cond}. 
This problem is generally well conditioned, because the equations are linear and the matrix $\mathbf{F}(t)$ is nilpotent.
We can thus employ standard Runge-Kutta-methods~\cite{Runge:1895hdo,Kutta:1901nid} from numerics to solve it.
These methods fundamentally work by discretizing the interval. At each point the values of the functions in the system of differential equations are estimated using the values at the previous point and their derivatives at some intermediate points.
The size of the discrete steps can be varied adaptively by some algorithms. We obtain in this way an effective method to reduce the numerical evaluation of iterated integrals to solving linear systems of differential equations. For this problem well-established numerical libraries exist, and we discuss our choices in section~\ref{sec:codes}. The only input needed are numerical routines to evaluate the letters $f_i(t)$. This is typically a much simpler problem. 

So far we have focused on convergent integrals that do not require regularisation. Let us now briefly discuss what happens in case $f_1(t)$ has a simple pole at $t=0$, so that we need to interpret the iterated integral as a shuffle-regulated version. The system of differential equations in eq.~\eqref{eq:diffEq} remains unchanged. The integrals, however, may develop logarithmic singularities in the limit $t\to 0$, as can be seen for example from the appearance of explicit logarithms in eq.~\eqref{eq:regularize}. After regularisation, we only need to solve differential equations for convergent integrals, whose initial conditions are given by the fact that convergent integrals vanish at $t=0$.

Let us discuss some optimisations that one may apply to this algorithm. 
First, regularisation may increase the amount of integrals that need to be evaluated, and therefore also the computational effort required.
To reduce this effect, we first note that from eqs.~\eqref{eq:hatIDefinition} and~\eqref{eq:regulator_def_eps} one can derive the same system of differential equations~\eqref{eq:system_main_matrix} also for a regulated iterated integral in a completely analogous fashion assuming non-zero $t$.
If we are given $\mathbf{I}(\tilde{t})$ for a fixed (small) $\tilde{t}\neq 0$, we can compute $\mathbf{I}$ at arbitrary points.
In this way we can avoid the massive increase in the number of differential equations caused by the regularisation procedure.
To turn this into a practical algorithm, we need to proceed in two steps.
First, we apply the regularisation to the iterated integrals $\textbf{I}(\tilde{t})$ and we evaluate them numerically using our code.
Secondly, we use the system of differential equations fulfilled by $\mathbf{I}(t)$ to compute the value of $I(\omega_1,\dots,\omega_n; t)$ at arbitrary $t$.
This procedure obviously does not avoid the regularisation procedure fully, but the regularised expression only needs to be evolved with the differential equation over the (much shorter) line segment $(0,\tilde{t})$.
Depending on the number of integrals produced by the regularisation, this may reduce the computational effort significantly.

A second optimisation stems from the fact that in applications to multi-loop Feynman integral or scattering amplitudes, one is typically not interested in evaluating just one iterated integral, but usually hundreds, if not thousands of them. It is therefore important to minimise the duplication of workflows and the number of quantities that need to be evaluated. In our method, we do not only evaluate a single iterated integral, but the solution to the linear system also provides all integrals of lower length obtained by stripping off some letters from the right. Different iterated integrals may of course reduce to the same lower-length integrals, and so it is natural to try to avoid that the same lower lengths integrals are evaluated multiple times.
As an example, consider the iterated integrals $I(\omega_1, \omega_2; t)$ and $I(\omega_1, \omega_3; t)$, where $\omega_i$ are some holomorphic one-forms.
Using the previously discussed strategies, we would solve two systems of differential equations:
\beq\bsp
\dv{t}\begin{pmatrix}I(\omega_1,\omega_2;t)\\
I(\omega_1;t)\\
I(;t)\end{pmatrix}&\,= 
\begin{pmatrix}0& f_2(t) & 0\\
0&0&f_1(t)\\0&0&0\end{pmatrix} \begin{pmatrix}I(\omega_1,\omega_2;t)\\
I(\omega_1;t)\\
I(;t)\end{pmatrix}\,,\\
\dv{t}\begin{pmatrix}I(\omega_1,\omega_3;t)\\
I(\omega_1;t)\\
I(;t)\end{pmatrix}&\,= 
\begin{pmatrix}0& f_3(t) & 0\\
0&0&f_1(t)\\0&0&0\end{pmatrix} \begin{pmatrix}I(\omega_1,\omega_3;t)\\
I(\omega_1;t)\\
I(;t)\end{pmatrix}\,.
\esp\eeq
As both systems contain $I(\omega_1; t)$, this integral is computed twice. This can be avoided by combining the two systems into a single larger system,
\beq
\dv{t}\begin{pmatrix}I(\omega_1,\omega_2;t)\\I(\omega_1,\omega_3;t)\\
I(\omega_1;t)\\
I(;t)\end{pmatrix}= 
\begin{pmatrix}0&0& f_2(t) & 0\\
0&0& f_3(t) & 0\\
0&0&0&f_1(t)\\0&0&0&0\end{pmatrix} \begin{pmatrix}I(\omega_1,\omega_2;t)\\I(\omega_1,\omega_3;t)\\
I(\omega_1;t)\\
I(;t)\end{pmatrix}\,.
\eeq
We see that in this way we can extract both $I(\omega_1,\omega_2;t)$ and $I(\omega_1,\omega_3;t)$ from the solution of a single system. Moreover, from this system we can also extract $I(\omega_1;t)$, so if also this integral appears in the expression for a Feynman integral, it does not need to be evaluated separately. 

\begin{figure}[t]
    \centering
    \includegraphics[width=0.8\textwidth]{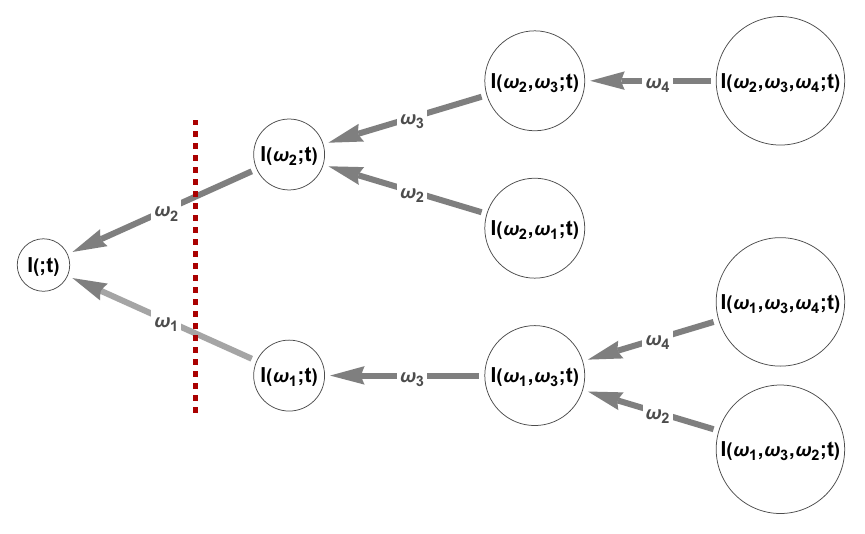}
    \caption{The resulting tree graph representing the relations between the iterated integrals $I(\omega_1,\omega_3,\omega_2;t)$, $I(\omega_1,\omega_4,\omega_4;t)$, $I(\omega_2,\omega_1;t)$ and $I(\omega_2,\omega_3,\omega_4; t)$. Each arrow represents taking a derivative. The edges passing through the dashed line are removed for the computation (see the discussion in the main text).}\label{fig:graph}
\end{figure}

The previous example can be turned into an effective algorithm also for more complicated cases.
We model the dependencies between the kernels in a set of iterated integrals by a tree graph. Each node of the tree represents an iterated integral, and its parent is the same integral with the last letter removed.
This tree is naturally rooted, and the root is given by the integral of length zero. An example of such a rooted tree constructed from three iterated integrals of length three and one integral of length two can be found in figure~\ref{fig:graph}.
To convert the tree into a system of differential equations, the derivative of each node is given by the value of the parent node times the previously dropped integration kernel.
We note, however, that if implemented naively, this procedure may have a significant downside in some situations. 
We can understand this on our example though figure~\ref{fig:graph}, if we duplicate $I(;t)=1$ the matrix describing the system can be brought into block-diagonal form
\begin{align}
    \mathbf{F}&=\mqty(0&0&f_4&0&0\\
    0&0&0&f_2&0\\
    0&0&0&f_3&0\\
    0&0&0&0&f_2\\
    0&0&0&0&0\\
    &&&&&0&0&f_4&0&0\\
    &&&&&0&0&f_2&0&0\\
    &&&&&0&0&0&f_3&0\\
    &&&&&0&0&0&0&f_1\\
    &&&&&0&0&0&0&0)\,,
\end{align}
were the top-left block corresponds to the iterated integrals starting with the letter $\omega_2$ and the bottom-right block to all iterated integrals starting with the letter $\omega_1$.
Instead of considering this as one large system, it is possible to split it into two independent, smaller systems.
Graphically, this corresponds to cutting the tree along the dashed line in figure~\ref{fig:graph}.
Afterwards the systems of differential equations are as small as possible without introducing any redundancies (except for $I(;t)$, which is equal to 1).
The numerical solvers can then integrate them quicker, because for smaller systems generally better choices for the step size can be made.
In this tree picture another small optimisation can be performed by only evaluating each of the $\omega_i$ once for every value of $t$ and caching the result.
Between different systems of differential equations this is not possible, as the exact values at which the $\omega_i$ are evaluated may differ.

% !TEX root = main.tex

\section{The \IterInt\ package}\label{sec:codes}
We have implemented the algorithms described in the previous section into a package called \IterInt, both in \mathematica\ and \cpp.
The main differences between the codes are due to different libraries and algorithms used to numerically solve the differential equations.
Generally, the algorithms used by \mathematica\ are more robust and flexible, while the \cpp\ code is significantly faster at reasonable precision. In the remainder of this section, we describe these implementations in turn, focusing on simple examples that illustrate their usage.
An example \mathematica\ notebook as well as \cpp\ source code are provided together with the package. The package is available on \textsc{GitHub}:

\vskip 2mm
    \url{https://github.com/baugid/IterInt}

\subsection{Details on the Mathematica implementation}\label{sec:mathematica}
We now describe the usage of the \mathematica\ package.
The package can simply be loaded using \texttt{Get}, if its folder is contained in \texttt{\$Path}:\footnote{Alternatively, the user may first use the \texttt{SetDirectory[]} function of \mathematica\ to set the current directory to the folder where the file \texttt{IterInt.m} file is located.}
\begin{verbatim}
    << IterInt`
\end{verbatim}
In the remainder of this section, we describe the main usage of the package, focusing on the definition of integration kernels, the numerical evaluation of iterated integrals and the regularisation of divergent integrals.

\paragraph{Definition of new integration kernels.} We start by illustrating on an example how to define new integration kernels. We focus on the kernels defined by the differential forms:
\beq\label{eq:example_kernels}
\omega_1 = \mathrm{d}\xi\,\cot \xi\,,\qquad\omega_2 = \mathrm{d}\xi\,\xi^2\,,\qquad \omega_3=\frac{\mathrm{d}\xi}{\xi}\,.
\eeq
These kernels can be defined in \IterInt\ by issuing the commands:
\begin{verbatim}
    defineKernel[cot, Cot, 1];
    defineKernel[square, #^2 &];
    defineKernel[inv, #^-1 &, 1];
\end{verbatim}
The first argument of \texttt{defineKernel} is the identifier, which is a symbol that will be used later to reference this integration kernel.\footnote{
    In principle, this can be an arbitrary expression, though it is highly advised for it to be a symbol.}
The second argument is a function that takes one argument and numerically evaluates the integration kernel. Here it does not matter if it is an internal function (like \texttt{Cot}), an anonymous function\footnote{Sometimes also called \emph{pure function} in \mathematica.} or a custom named function.
The value of the residue at the origin can be passed as an optional third argument. If it is omitted, then the default value is assumed to be $0$.
For functions that do not have a pole at the origin, it is recommended for the residue to be identical to 0 (i.e., it should be the integer 0, rather than a floating point approximation).

It is possible to obtain a list of all defined kernels using \texttt{listKernels}. The definition of a kernel is returned by \texttt{exportKernel}, which returns an association containing the name, functional expression, and the value of the residue at the origin.
In our example, the output of \texttt{exportKernel[cot]} is
\begin{verbatim}
    <|name -> cot, expression -> Cot, residue -> 1|>
\end{verbatim}
Note that \texttt{defineKernel} supports such an association as its sole argument to reimport a kernel.
A list of all methods for working with the kernels is provided in table~\ref{tab:kernelMethods}.
Some further information can be obtained by using the \texttt{Information[]} command, which outputs the usage messages in \mathematica.

\bgroup
\def\arraystretch{1.25}
\begin{table}[tb]
    \begin{tabularx}{\linewidth}{cX}
        \toprule
        Signature                                  & \multicolumn{1}{c}{Explanation}                                                                                             \\
        \midrule
        \texttt{defineKernel[name, func, residue]} & Defines a new kernel referenced by \texttt{name} with value \texttt{func[t]} at $t$ and (optionally) given residue in $0$.  \\
        \texttt{listKernels[]}                     & Lists all defined kernels by their name.                                                                                    \\
        \texttt{exportKernel[name]}                & Returns an association describing the kernel referenced by \texttt{name}. The output is supported by \texttt{defineKernel}. \\
        \texttt{exportKernels[]}                   & Returns a list of the output of \texttt{exportKernel} for all defined kernels.                                              \\
        \texttt{deleteKernel[name]}                & Deletes the kernel referenced by \texttt{name}.                                                                             \\
        \bottomrule
    \end{tabularx}
    \caption{List of commands to manipulate and define kernels.}\label{tab:kernelMethods}
\end{table}
\egroup

\paragraph{Numerical evaluation of iterated integrals.} We now describe how we can numerically evaluate iterated integrals built from the kernels declared using the \texttt{defineKernel} function. The integral $I(\omega_1,\dots,\omega_n;t)$ is represented inside the \IterInt\ package by an object of the form \texttt{IIntegral[f1,\dots,fn,t]}, where the \texttt{f}$i$ refer to the identifiers of the declared integration kernels. For example, for the kernels defined in eq.~\eqref{eq:example_kernels}, the iterated integral $I(\omega_1,\omega_2,\omega_3;t)$ is represented by the object \texttt{IIntegral[cot, square, inv, t]}. Note that we always assume that the path defining the iterated integrals is the straight line from $0$ to $t$.

Consider now a \mathematica\ expression \texttt{expr} involving \texttt{IIntegral} objects where the variable $t$ takes some numerical value. For example, we may consider
\begin{verbatim}
    expr = IIntegral[cot, square, inv, 3/2];
\end{verbatim}
The expression \texttt{expr} can be arbitrarily complicated (as a long as it is a valid \mathematica\ expression).
To evaluate the iterated integrals appearing in \texttt{expr}, it suffices to issue the command
\begin{verbatim}
    computeIIntegrals[expr]
\end{verbatim}
As a second optional argument, a list of rules can be passed that will be applied to \texttt{expr} before numerical evaluation.
\IterInt\ then evaluates all objects with head \texttt{IIntegral} in \texttt{expr} numerically (after regularising the iterated integrals, if required), and returns a version of expression with these numerical values inserted.\footnote{During this process, if the forms have poles at the origin, the way \texttt{NDSolve} works, some intermediate expressions may occur which are infinite, and \mathematica\ may emit the messages \texttt{Power::infy} and \texttt{Infinity::indet}.
    They can safely be ignored or switched off. For this reason, these messages have been turned off for the time of evaluation with \texttt{computeIIntegrals}.}
To steer the evaluation, various different options can be provided.
A large part of them are inherited from the underlying \texttt{NDSolveValue}, which are just passed through.
The desired precision and accuracy can then be requested by means of \texttt{WorkingPrecision}, \texttt{AccuracyGoal} and \texttt{PrecisionGoal}.
In our example, we can obtain a result with higher precision via
\begin{verbatim}
    computeIIntegrals[expr, WorkingPrecision -> 50, PrecisionGoal -> 30]
\end{verbatim}
In addition to the options of \texttt{NDSolve}, various other options (specific to \IterInt) are available, which we now describe in some detail.
A complete list of additional options can be found in table~\ref{tab:computeOptions}.
In particular, the option \texttt{integrationMethod} determines the internal solving algorithm used.
The possible values and their effects are the following:

\begin{itemize}
    \item \underline{\texttt{treeIntegrate}:}
          This is our default method, which evaluates all iterated integrals in \texttt{expr} at once by constructing a rooted tree from them in order to avoid redundancies in the workflow. This approach is also useful if the total number of iterated integrals is low, but after the regularisation many iterated integrals occur and the resulting integrals are highly related and thus significantly profit from being combined.
    \item\underline{\texttt{plainIntegrate}:} This method allows one to disable the construction of the rooted tree encoding the interrelations between iterated integral, and all iterated integral are evaluated independently from each other.
    \item \underline{\texttt{splittingPlainIntegrate}:} This method is based on \texttt{plainIntegrate} and only evaluates the systems of differential equations produced by regularisation on a short path. For the remaining distance only the system involving the target integral itself is solved. In some cases this can reduce the runtime, if many convergent integrals are produced by the regularisation procedure. The point $\tilde{t}$ up to which the regulated integrals are computed is passed by the option \texttt{switchPoint} (see the discussion in section~\ref{sec:algo}).
          Note that since all operations are mapped to the domain $[0,1]$, the value of \texttt{switchPoint} necessarily needs to be between $0$ and $1$, regardless of the value of $t$.
    \item \underline{\texttt{splittingTreeIntegrate}:} The general structure of this algorithm is very similar to \texttt{splittingPlainIntegrate} (in particular the option \texttt{switchPoint} is also available here), but all evaluations of integrals are combined and performed using the tree-based approach. This combines the benefits from both optimisations. It is therefore useful in situations with many related iterated integrals, of which a significant portion requires regularisation. This is very likely to arise in the context of Feynman integrals.
\end{itemize}

\bgroup
\def\arraystretch{1.25}
\begin{table}[tb]
    \begin{tabularx}{\linewidth}{ccX}
        \toprule
        Option                     & Default                & \multicolumn{1}{c}{Description}                                                                                      \\
        \midrule
        \texttt{integrationMethod} & \texttt{treeIntegrate} & Algorithm to use for the computation.                                                                                \\
        \texttt{startRegulator}    & 0                      & Technical cutoff between $0$ and $1$ for numerical instabilities around the origin.                                      \\
        \texttt{regulator}         & 1                      & Value of $v\in\C^\times$ from eq.~\eqref{eq:regularize}.                                                                    \\
        \texttt{switchPoint}       & $2\times 10^{-5}$      & Value of $0<\tilde{t}\leq1$ for the splitting of the integration path.                                                               \\
        \texttt{Quiet}             & \texttt{True}          & If set to \texttt{False}, the messages \texttt{Power::infy} and \texttt{Infinity::indet} are not suppressed any more. \\
        %       \texttt{checkFunction}     & \texttt{(False \&)}       & Function that handles trivial integrals.                                        \\
        \bottomrule
    \end{tabularx}
    \caption{List of options for \texttt{computeIIntegrals}}\label{tab:computeOptions}
\end{table}
\egroup

Note that, in case integrable divergences occur at the origin, it is sometimes numerically required to start the integration not in 0, but (very) close to it.
This can be achieved using the option \texttt{startRegulator}. If it is set to a non-zero value $0<\alpha\ll 1$, the integration is started at $\alpha t$ instead of $0$, while the regularisation is performed as usual.
Note that a rough upper-bound for $\alpha$ is the required accuracy, since no attempt is made to recover the contribution along the segment $0$ to $\alpha t$.

\paragraph{Regularisation.} In case some of the integration kernels have simple poles at the origin, the iterated integrals may need to be regulated before a numerical evaluation can be attempted. \IterInt\ implements the shuffle regularisation reviewed in section~\ref{sec:regularisation}, and the function \texttt{computeIIntegrals} automatically performs the shuffle regularisation required for all kernels for which the user has provided a non-zero value for the residue at the origin.

In some cases, knowing the regulated expression analytically may be of independent interest, for example if one wants to extract the logarithmic singularities and the leading-power behaviour in the limit $t\to0$. \IterInt\ allows the user to output the regulated expression stand-alone, independently of the subsequent numerical evaluation. We illustrate this feature on the iterated integral
$I(\omega_3,\omega_1,\omega_2;t)$ (the one-forms $\omega_i$ have been defined in eq.~\eqref{eq:example_kernels}).
Since both $\omega_1$ and $\omega_3$ have a simple pole at the origin, the naive iterated integral is divergent, and we need to interpret $I(\omega_3,\omega_1,\omega_2;t)$ as a shuffle-regulated expression. Equation~\eqref{eq:regularize} gives
\begin{equation}
    \begin{aligned}
        \operatorname{Reg}_v\left[I(\omega_3,\omega_1,\omega_2;t)\right] & = I(\omega_2,\omega_1^\infty,\omega_3^\infty;t)-I(\overline{\omega}_1,\omega_2,\omega_3^\infty;t)-I(\overline{\omega}_1\omega_3^\infty,\omega_2;t) \\
                                                                         & +I(\overline{\omega}_3,\omega_1,\omega_2;t)+\qty{I(\overline{\omega}_1,\omega_2;t)-I(\omega_2,\omega_1^\infty;t)}\log\tfrac{t}{v}                  \\
                                                                         & +\inv{2}I(\omega_2;t)\log^2\tfrac{t}{v}\fullstop
    \end{aligned}
\end{equation}
The user can obtain this expression from within \IterInt\ by issuing the command
\begin{verbatim}
    regulateExpression[IIntegral[inv, cot, sq, t]]
\end{verbatim}
A second optional argument can be passed in to choose the regulator $v$ from eq.~\eqref{eq:regularize}. The output involves for example the integral $I(\overline{\omega}_1,\omega_2,\omega_3^{\infty};t)$, which is represented in \IterInt\ as
\begin{verbatim}
    IIntegral[regulated[cot], sq, pole[inv], t]
\end{verbatim}
The previous expression involves the kernels \texttt{regulated[cot]} and \texttt{pole[inv]}, which correspond to $\overline{\omega}_1$ and $\omega_3^{\infty}$ respectively.
We note that the kernels wrapped into \texttt{regulated} are computed internally by subtracting the singularity at the origin.  In some cases this may lead to numerical instabilities. The user may then want to provide an alternative expression for the regulated kernel that allows for a more stable numerical evaluation.
In the example above, this might look like\footnote{In this special case such a rewriting is not necessary to obtain a stable result.}
\begin{verbatim}
    kernelExpression[regulated[cot]] = (# Cos[#] - Sin[#])/(# Sin[#]) &
\end{verbatim}

The number of terms in the regulated expression can be rather large, owing to a swell of terms in the combinatorial definition of the regularisation in eq.~\eqref{eq:regularize}. In such cases,
even if the regularised expression in analytic form is not required per se, it can be useful to manipulate it before numerical evaluation, because this may speed the code substantially.
In our example, some of the integrals vanish, because $\overline{\omega}_3=0$.
Another situation where it might come in handy to manipulate the expression after regularisation is if some integrals are straightforward to compute and do not require involved numerics.
Such an integral would be
\begin{verbatim}
    IIntegral[sq, pole[cot], pole[inv], 3/2]
\end{verbatim}
from the same example above. All integrations necessary only involve monomials allowing to easily get the precise result $\tfrac{1}{8}$.
In that way four of the seven convergent integrals in our example can easily be dealt with.
For more involved cases it is also possible that some of the integrals cancel after regularisation (e.g., because some of the kernels $\overline{\omega}_i$ or $\omega_i^{\infty}$ are dependent).

\subsection{Details on the \texorpdfstring{\cpp}{C++} implementation}
\label{sec:cpp}
We now describe the \cpp\ implementation of \IterInt.
We will explain the usage of the \cpp\ code on the same example as in section~\ref{sec:mathematica}, focusing mostly on the differences to the \mathematica\ package.

\paragraph{Dependencies on external libraries.}
The \cpp\ implementation of \IterInt\ relies on external libraries, like the \boost\ library and GNU Scientific Library (\GSL)~\cite{BoostPage,GSLPage}, to solve ordinary differential equations and to handle high precision floating point arithmetic. We now briefly describe these dependencies.

We provide two different integrator backends to solve the differential equations, both of which are commonly used libraries in \cpp.
The first library is \emph{Boost.Numeric.Odeint} from the extensive family of \boost\ libraries. It is highly customizable and also allows calculations with arbitrary precision.
This may also result in relatively high compilation times.
As a second backend we provide an interface to \GSL, which is a large, efficient and well-tested C library for numerical computations.
It is not as flexible, and only allows for calculations with double precision.
As a wrapper around arbitrary precision arithmetic libraries like GNU MPFR and MPC~\cite{mpfr,mpc}, we use \textsc{Boost.Multiprecision}~\cite{BoostMultiprecDoc} for our examples.
Both our code and \emph{Boost.Numeric.Odeint} allow the user to easily choose among different libraries.

\paragraph{Including the \IterInt\ library.}

Our library is entirely contained in three header files. To use the \boost\ backend, one needs to include \texttt{integratorBoost.hpp}.
For the \GSL\ backend the relevant header file is \texttt{integratorGSL.hpp}. The third header \texttt{integratorBase.hpp} contains all the code common between the two backends and is only necessary to include directly if a custom backend is to be used.
It is also possible to use multiple backends in the same file.
To use GSL, the code has to be linked against \texttt{libgsl.so}.
Note that features from \cpp 20 are used. A reasonably modern compiler is thus required.

For now, we will focus on the \GSL\ backend, and we will come back to the differences with respect to \boost\ later.
To load our code with the \GSL\ backend one needs to use:
\begin{verbatim}
    #include "IterInt/integratorGSL.hpp"
\end{verbatim}
All added functionalities live in the namespace \texttt{iteratedIntegrals}. To keep the examples concise, we will assume the following using-directives:
\begin{verbatim}
    using namespace iteratedIntegrals;
    using namespace std;
    using cmplx = complex<double>;
\end{verbatim}

\paragraph{Definition of new integration kernels.}
Just like for the \mathematica\ implementation, the integration kernels have to be defined by the user. In the following we illustrate this again on the three kernels defined in eq.~\eqref{eq:example_kernels}. These can be defined as instances of the \texttt{Integrand} class:
\begin{verbatim}
Integrand cot{1,
  [](const cmplx& x){if(x == 0.) return cmplx{}; else return 1./tan(x);}};
Integrand sq{[](const cmplx& x){return x * x;}};
Integrand inv{1,
  [](const cmplx& x){if(x == 0.) return cmplx{}; else return 1./x;}};
\end{verbatim}
The template parameter of the \texttt{Integrand} class is the type used for complex numbers. The default is \texttt{complex<double>}. The first parameter is
the value of the residue at the origin, which is again passed as an optional parameter. The second parameter must be an object with defined \texttt{operator()} taking a constant reference to the type used for complex numbers as a parameter and returning an instance of the complex number type.\footnote{To be precise: It must be possible to construct a \texttt{std::function<cmplx(const cmplx\&)>} from the object passed in.}
This can be realised by using a lambda expression as done above.
At variance with the \mathematica\ implementation, it is required to always return a finite value. This also holds at the origin for functions that have non-vanishing residue there. In our examples, such functions return zero if the argument is zero.
We note that it can happen that the function objects are copied multiple times during a calculation.
To keep the memory footprint and runtime small, it is recommended to ensure that the function objects are cheap to copy.

\paragraph{Numerical evaluation of iterated integrals.}
Let us now discuss how we can evaluate iterated integrals using the GSL backend.
This is achieved by an instance of \texttt{GSLIntegrator}.
Its constructor parameters fundamentally have the same purpose as the options of the \texttt{computeIIntegrals} function in \mathematica, and their default values are given in table~\ref{tab:GSLargs}.
The precision and accuracy (corresponding to \texttt{PrecisionGoal} and \texttt{AccuracyGoal} in \mathematica) are controlled by the two parameters $\varepsilon_\text{rel}$ and $\varepsilon_\text{abs}$.
They are used to define an upper bound for the error estimate $\delta y$ of the approximate solution $y(t)$ of the system of differential equations.
This upper-bound is given by~\cite{BoostOdeintDoc}
\begin{align}\label{eq:OptimizationTarget}
    \abs{\delta y} & \leq \varepsilon_\text{abs}+\varepsilon_\text{rel} \qty(\abs{y}+\abs{\derivative{y}{t}})\fullstop
\end{align}
Generally, it is not possible to set either $\varepsilon_\text{abs}$ or $\varepsilon_\text{rel}$ to zero, as otherwise the solver may become unstable.
If the error estimate exceeds the requested bound, the step must be rejected and repeated with a smaller step size $h$.
It is known (cf., e.g., refs.~\cite{bartels2025numerical,hanke2006grundlagen}) that for a given algorithm, we have
\begin{align}
    \abs{\delta y} & \propto h^{-n}\comma
\end{align}
where $n$ is a constant that only depends on the algorithm used. This allows one to compute an estimate for the step size where equality in eq.~\eqref{eq:OptimizationTarget} holds.
In practice, some measures to prevent excessive increases and decreases of $h$ are implemented. For example, the maximum increase/decrease in one step is a factor of $5$. We refer to the documentation of the libraries~\cite{BoostOdeintDoc,GSL} for more details.
To initialize the solver, a starting step size $h_0$ needs to be chosen. In most cases, the effect of this choice on runtime and accuracy should be minute, as long as the error estimate around the origin is reasonably accurate.
If the error estimate is too small, it might become necessary to limit the step size from above. This can be achieved by the parameter $h_\text{max}$.
The remaining parameters are identical to the \mathematica\ implementation discussed in section~\ref{sec:mathematica}.
In most applications, only the precision needs to be adapted. For example, the code to construct an integrator with $\varepsilon_\text{abs}=\num{1e-12}$ and $\varepsilon_\text{rel}=\num{1e-17}$ would be:
\begin{verbatim}
    GSLIntegrator integrator(1e-12, 1e-17);
\end{verbatim}

\bgroup
\def\arraystretch{1.25}
\begin{table}[tb]
    \begin{tabularx}{\linewidth}{ccX}
        \toprule
        Parameter                & Default                      & \multicolumn{1}{c}{Description}                                                                                                                 \\
        \midrule
        $\varepsilon_\text{abs}$ & \num{1e-12}                  & Absolute precision goal, see eq.~\eqref{eq:OptimizationTarget}.                                                                      \\
        $\varepsilon_\text{rel}$ & \num{1e-12}                  & Relative precision goal, see eq.~\eqref{eq:OptimizationTarget}.                                                                      \\
        $h_0$                    & \num{1e-5}                   & Initial step size of the numerical integration method.                                                                                            \\
        $h_\text{max}$           & 1                            & Maximal step size of the numerical integration method. $h_\text{max}\geq 1$ corresponds to no limit.                                              \\
        \texttt{startRegulator}  & 0                            & Technical cutoff between $0$ and $1$ for numerical instabilities around the origin.                                                                 \\
        \texttt{switchPoint}     & \num{2e-5}                   & Value of $0<\tilde{t}\leq1$ for splitting integration.                                                                                          \\
        \texttt{gslAlgorithm}    & \verb|gsl_odeiv2_step_rk8pd| & Algorithm to use provided by \GSL. Possible values can be found in the documentation~\cite{GSL}. Methods requiring the Jacobian cannot be used. \\
        \texttt{regulator}       & 1                            & Value of $v\in\C^\times$ from eq.~\eqref{eq:regularize}.                                                                                 \\
        \bottomrule
    \end{tabularx}
    \caption{List of constructor parameters for \texttt{GSLIntegrator} in order.}\label{tab:GSLargs}
\end{table}
\egroup

To evaluate iterated integrals, the chosen member method depends on the chosen strategy. The names of the different methods are identical to those used in the \mathematica\ implementation.
A single integral can be computed using \texttt{plainIntegrate}. For example, the following code returns the numerical value of the iterated integral $I(\omega_1,\omega_2,\omega_3;\tfrac{3}{2})$ (with $\omega_i$ defined in eq.~\eqref{eq:example_kernels}):
\begin{verbatim}
    cmplx res = integrator.plainIntegrate(
      cmplx{3/2., 0},
      {cot, sq, inv}
    );
\end{verbatim}
The integration limits are passed as the first argument, and the second argument is a \texttt{vector} of \texttt{Integrand} objects. Combined, these two arguments define the iterated integral to be evaluated.
We can also evaluate multiple iterated integrals at once, and also speed up the evaluation using the tree-based approach, using the member function \texttt{treeIntegrate}. For example, the following code evaluates both $I(\omega_1,\omega_2,\omega_3;\tfrac{3}{2})$ and $I(\omega_3,\omega_1,\omega_2;\tfrac{3}{2})$.
\begin{verbatim}
    auto resList = integrator.treeIntegrate(
      cmplx{3/2., 0},
      {{&cot, &sq, &inv}, {&cot, &inv, &sq}}
    );
\end{verbatim}
Note that this integral requires regularisation, which is done automatically by the code.
As multiple integrals need to be passed in, the second argument changes to a \texttt{vector} of \texttt{vector} objects.
Also, to allow our code to identify identical integrands, the value type changes to pointers to \texttt{Integrand} objects.
The result is in the form of a \texttt{vector} as well.

Unlike for the \mathematica\ package, we do not provide an interface to manipulate the expanded expression after the regularisation. Nevertheless, it is still possible to manually provide values of simple iterated integrals.
For this a third optional argument to all methods is available. Skipping all the trivial iterated integrals in the above example is achieved by the following code:
\begin{verbatim}
    resList = integrator.treeIntegrate(
      cmplx{3/2., 0},
      {{&cot, &sq, &inv}, {&cot, &inv, &sq}},
      [&inv](auto integ){
        if(integ[0].isRegulated() && integ[0].fct == &inv)
          return optional(cmplx{});
        return optional<cmplx>{};
      });
\end{verbatim}
This optional argument is a function, which takes a \texttt{span} of so-called \texttt{TaggedFunction} objects describing an iterated integral as a parameter.
A \texttt{TaggedFunction} has the same purpose as the \texttt{regulated} and \texttt{pole} wrappers in \mathematica.
These properties can be checked using the member functions \texttt{isRegulated} and \texttt{isPole} respectively.
The underlying pointer to an \texttt{Integrand} can be obtained using the \texttt{fct} member.
In this example, if the first integration kernel is given by the regulated version of $\omega_3=\tfrac{\mathrm{d}\xi}{\xi}$, an \texttt{optional} containing a value (namely $0$) is returned.
Otherwise, the value is not known, so an empty \texttt{optional} is returned.

We note that the two additional  member functions \texttt{splittingPlainIntegrate} and \texttt{splittingTreeIntegrate} are also available.
They have the same signature as their discussed counterparts and follow the different regularisation strategy as discussed for the \mathematica\ implementation.

\paragraph{The \boost\ backend.}
We now discuss the differences to the \boost\ backend.
First, the header file \texttt{integratorBoost.hpp} needs to be included. Also the construction of a \texttt{BoostIntegrator} is slightly different, as we now explain.

The class \texttt{BoostIntegrator} is templated with three type parameters.
The first two are the real and complex data types, respectively. This allows one to work with arbitrary precision calculations by choosing such types, e.g., from \textsc{Boost.Multiprecision}.
If no choice is made, the defaults are \texttt{double} and \texttt{complex<double>}.
The third parameter selects the solver used to solve the differential equations.
By default, we use the \verb|runge_kutta_fehlberg78| implementation.
To avoid (unreadable) compiler errors, we note a couple of requirements of the chosen types.
The two chosen number types must be compatible in the sense that the real type can be constructed from the real and imaginary parts of the complex type.
The complex type needs to have a very similar syntax to the standard library type \texttt{complex}.
Any  chosen solver must obviously work with the corresponding number types. In addition, it is necessary to be compatible with \texttt{boost::odeint::make\_controlled}.
Available methods from \boost\ can be found in the documentation~\cite{BoostOdeintDoc}, and they have template parameters themselves. In the naming conventions of \boost, the \texttt{State} must be \texttt{vector<cmplx>}, where \texttt{cmplx} is the chosen type for complex numbers, and \texttt{Value} needs to match the real number type.
Further template parameters are generally correctly resolved automatically.
In practice, for double precision only the name of the class changes, and the constructor parameter \texttt{gslAlgorithm} is removed.

Let us now repeat the same calculation as with \GSL\ using \boost\ to $30$-digits of precision.
We choose the arbitrary precision types from \textsc{Boost.Multiprecision}, and we include the appropriate headers:
\begin{verbatim}
    #include <boost/multiprecision/mpc.hpp>
    #include <boost/multiprecision/mpfr.hpp>
\end{verbatim}
As previously discussed, we use \boost\ as a wrapper around \textsc{Mpc} and \textsc{Mpfr}.
Note that the code then needs to be linked against the corresponding libraries.
To keep the code compact, we use using-directives similar to the \GSL\ example:
\begin{verbatim}
    using nr = boost::multiprecision::mpfr_float;
    using cmplx = boost::multiprecision::mpc_complex;
    using namespace std;
    using namespace iteratedIntegrals;
\end{verbatim}
To set up the arbitrary precision types, we need to specify a default precision:
\begin{verbatim}
    nr::default_precision(50);
    cmplx::default_precision(50);
\end{verbatim}
Both precisions should generally be kept the same to avoid precision losses.
With this setup and the above discussion in mind, we define the integrator
\begin{verbatim}
    BoostIntegrator<nr, cmplx> integrator(1e-40, 1e-30);
\end{verbatim}
To make use of the precision available, we set the precision target to $\num{1e-40}$ (absolute) and $\num{1e-30}$ (relative).
The definition of the integration kernels is essentially identical to what was discussed before, with the exception that we have to be a bit more verbose with the types, because we are using custom data types and want to ensure sufficient precision at all times:
\begin{verbatim}
    Integrand<cmplx> cot{1,
      [](const cmplx& x){
        if(x == 0.) return cmplx{}; else return cmplx{1}/tan(x);}};
    Integrand<cmplx> sq{[](const cmplx& x){return x * x;}};
    Integrand<cmplx> inv{1,
      [](const cmplx& x){
        if(x == 0.) return cmplx{}; else return cmplx{1}/x;}};
\end{verbatim}

Once the integration kernels have been defined in this way, the numerical evaluation of the iterated integrals is identical to the \GSL\ case, and so we do not show examples here for the \boost\ backend.
We note, however, that in general it is necessary that the integration bound is provided with sufficient accuracy as well.
How this can be achieved depends an the type used and can be found in the relevant documentation, cf., e.g., refs.~\cite{BoostMultiprecDoc,mpc,mpfr}.

\paragraph{Calling the \cpp\ code from within \mathematica.}
It is possible to use our \cpp\ code from within \mathematica\ in a slightly restricted way.
For this a library file called \texttt{libIntegrate} (with the correct file ending), which is provided by us for Linux systems, must be present in the same folder as \texttt{IterInt.m}.
To use this with other operating systems the corresponding source code is provided as \texttt{mathematicaLink.cpp} and must be compiled and linked against \GSL.
Additionally, before loading the library, the flag \verb|$NativeIteratedIntegral| must be set to \texttt{True}.
Afterwards, this backend can simply be accessed by replacing \texttt{computeIIntegrals} with \texttt{computeIIntegralsC}.
The options steering the integration are then the same as for the \cpp\ code.
Note, however, that no arbitrary precision support exists and a significant performance overhead might occur, if the kernel expressions are not \texttt{ForeignCallback} objects.
% !TEX root = main.tex

\section{Examples: Ordinary and elliptic multiple polylogarithms}\label{sec:examples_MPLs}

In this section we illustrate the usage of \IterInt\ to evaluate numerically two well-established classes of iterated integrals, namely ordinary and elliptic multiple polylogarithms. There are public libraries for their numerical evaluation~\cite{Gehrmann:2001pz,Gehrmann:2001jv,Vollinga:2004sn,Buehler:2011ev,Frellesvig:2016lxm,Naterop:2019xaf,Walden:2020odh,Duhr:2026ell}. We stress that these libraries (with the exception of ref.~\cite{Walden:2020odh}) are specialised codes aiming to evaluate specific classes of special functions. This offers the possibility to design and optimise the corresponding numerical routines. \IterInt, instead, aims at the evaluation of general classes of iterated integrals, and is agnostic about the form of the integration kernels. Nevertheless, a direct comparison between \IterInt\ and existing libraries to evaluate (elliptic) polylogarithms is interesting as it gives us a way to assess the overall performance and the expected runtimes of our code.

\subsection{Multiple polylogarithms}
Let us start by discussing the numerical evaluation of multiple polylogarithms (MPLs), defined as the iterated integrals~\cite{LappoDanilevsky1935,Goncharov:1998kja,Remiddi:1999ew}:
\begin{align}\label{eq:MPL_def}
    G(a_1,\dots,a_n; z) & \coloneqq \int_0^z \dd{z_1} \inv{z_1-a_1}G(a_2,\dots,a_n; z_1)\fullstop
\end{align}
In the notations of section~\ref{sec:iterated}, MPLs correspond to
\beq
G(a_1,\dots,a_n; z) = I(\omega_n,\ldots,\omega_1;z)\,, \textrm{~~~with~~~} \omega_i = \frac{\dd{t}}{t-a_i}\,.
\eeq
To regularise these integrals, it is sufficient to set
\begin{align}
    G(\underbrace{0,\dots,0}_\text{$n$-times}; z) & \coloneqq \inv{n!}\log^n z\fullstop
\end{align}
Note that this is consistent with the shuffle-regularisation discussed in section~\ref{sec:iterated}.
Algorithms to evaluate MPLs are available, and they have been implemented into public packages~\cite{Gehrmann:2001pz,Gehrmann:2001jv,Vollinga:2004sn,Buehler:2011ev,Frellesvig:2016lxm,Naterop:2019xaf}.

In the following we focus on the algorithms from ref.~\cite{Vollinga:2004sn} implemented into \ginac~\cite{Bauer:2000cp}, because it allows one to handle the numerical evaluation of MPLs for arbitrary values of the arguments and with arbitrary precision.
We will perform numerical evaluations of a selected set of MPLs. We use a benchmark evaluation with \ginac\ at very high precision, which we consider the `true' value, and we request \IterInt\ to achieve a target deviation of less than $10^{-12}$, and less than $10^{-30}$ where possible.
In this way we can get an estimate of the comparative scalings of the actual precision and the runtimes. For the \mathematica\ implementation of \IterInt, instead of requesting a target deviation of less than $10^{-12}$, no options are passed to the solver in \mathematica\ in order to get an accurate representation of the default behaviour of the code.
For the calculations at high precision, we use a floating point accuracy of 70 decimal digits in \cpp\ and 60 digits in \mathematica.
Except when stated otherwise, we use \texttt{plainIntegrate} as the integration method, which provides a baseline for the other methods.

\begin{table}[tb]
    \centering
    \begin{tabular}{c c c c}
        \toprule
        Library      & Real part                                & Imaginary part                           & Runtime                 \\
        \midrule
        \ginac       & 0.076182801138147\ldots                  & -0.020626439348649\ldots                 & \SI{30}{ms}             \\
        \GSL         & 0.0761828011{\color{red}24328\ldots}     & -0.0206264393{\color{red}52445\ldots}    & \SI{435}{\micro\second} \\
        \boost       & 0.076182{\color{red}723485922\ldots}     & -0.0206264\color{red}07102772\ldots      & \SI{364}{\micro\second} \\
        \mathematica & 0.07618\color{red}3497939829\ldots       & -0.020626\color{red}291724825\ldots      & \SI{33}{ms}             \\
        \midrule
        \midrule
        \ginac       & \ldots75295691486784784\ldots            & \ldots96223263483207115\ldots            & \SI{54}{ms}             \\
        \boost       & \ldots7529569\color{red}0712835612\ldots & \ldots96223263\color{red}159376397\ldots & \SI{3.43}{\second}      \\
        \mathematica & \ldots7529569148678\color{red}3789\ldots & \ldots9622326348320\color{red}9149\ldots & \SI{2.28}{\second}      \\
        \bottomrule
    \end{tabular}
    \caption{Numerical evaluation of the MPL in eq.~\eqref{eq:MPL_ex_1} using \ginac\ as well as the \mathematica\ implementation and the \GSL\ and \boost\ \cpp\ backends of \IterInt. Digits in red are incorrect. Below the double line are the results at high precision, and the decimal expansion shown starts with the last digit of the \ginac\ result above.}\label{tab:firstGres}
\end{table}

As a first example, we compare the codes for the MPL
\beq\label{eq:MPL_ex_1}
G(1, 5, 3, 0, 0; 1 + i)\approx 0.0762\ldots -i\,0.0206 \ldots\,.
\eeq
Note that this particular MPL requires shuffle regularisation, as the right-most letter in eq.~\eqref{eq:MPL_ex_1} is a 0.
The results of the different codes are presented in table~\ref{tab:firstGres}.
We can see that the \cpp\ implementations of \IterInt\ are significantly faster than the rest at low precision.
At the same time, the precision of the \boost\ implementation is quite low.
This can be traced back to the fact that the algorithms in the \boost\ library are not able to cleanly deal with systems of differential equations only valid for $t>0$ and where the limit $t\rightarrow 0$ yields non-vanishing derivatives.
As the error depends on the initial step size $h_0$, the problem can be reduced by decreasing the initial step size to the wanted precision. The amount of additional steps due to this grows like $-\log_5(h_0)$.
For this purpose, we choose an initial step size of \num{1e-10} at low precision and \num{1e-20} at high precision in the following whenever necessary.
The accuracy then reaches the level of \GSL\ with a runtime of \SI{550}{\micro\second}.
\ginac\ reaches a deviation of less than \num{1e-16}, even though this accuracy was not requested.
The \mathematica\ code is notably the slowest with the worst precision. This is fundamentally caused by the low default values for \texttt{AccuracyGoal} and \texttt{PrecisionGoal}.
Mitigating this is possible at a small runtime penalty.
At high precision \ginac\ is significantly faster than \IterInt. This can be traced back to the difference in methodology of how the numerical evaluation is performed (see the discussion below).
\mathematica\ is faster than \boost\ as well, which hints at a better solving algorithm used.

The previous example is already quite exemplary, as it illustrates what we expect to be a general point.
For high-precision evaluations of MPLs, a specialised method like the one implemented into \ginac\ should generally be preferred over the generic approach provided by \IterInt.
This becomes obvious when the scaling of the error terms is compared. The algorithms implemented into \ginac\ are based on series expansions, and for a series expansion truncated after $n$ terms the error usually behaves like $\mathcal{O}(x^n)$, where $x$ is the (small) expansion parameter.
For numerical ODE solvers making $n$ steps, the error is $\mathcal{O}(n^{-p})$, for some exponent $p$ depending on the details of the method. We therefore generically expect a better convergence from series expansions than from ODE solvers (provided that the expansion parameter is small enough).

We can also use this specific example to test the effects of changing some of the settings used in the algorithms.
Using the \texttt{splittingPlainIntegrate} method, the low precision evaluation with both \cpp\ backends needs about \SI{30}{\percent} less time while the accuracy remains almost unchanged. The MPL in eq.~\eqref{eq:MPL_ex_1} needs to be regulated, and after regularisation we have to evaluate numerically several convergent integrals.
 We can speed up the evaluation using the tree-based algorithm and dropping trivial iterated integrals increases performance further (though the effect of the latter is small in the present case).
Without changing the resulting accuracy, the runtime reduces to about half of the basic approach.
Moreover, we can draw some interesting conclusions on the \boost\ backend that will continue to show throughout the other examples: as long as double precision is sufficient, \GSL\ is often as quick as \boost, but with better and more stable results.
We also observe that at high precision the \boost\ algorithms can sometimes underestimate the error.

\begin{table}[tb]
    \centering
    \begin{tabular}{c c c c}
        \toprule
        Library      & Real part                                & Imaginary part                           & Runtime                 \\
        \midrule
        \ginac       & 0.004706043557637\ldots                  & \ldots7998090133\color{red}604085\ldots  & \SI{8}{ms}              \\
        \GSL         & 0.004706043557\color{red}711\ldots       & \ldots79980901\color{red}87408323\ldots  & \SI{38}{\micro\second}  \\
        \boost       & 0.00470604355\color{red}8306\ldots       & \ldots799809\color{red}1014963176\ldots  & \SI{50}{\micro\second}  \\
        \mathematica & 0.004706\color{red}217804609\ldots       & \ldots79\color{red}43015348512048\ldots  & \SI{6}{ms}              \\
        \midrule
        \midrule
        \ginac       & \ldots79300639239059010\ldots            & \ldots92144019201977752\ldots            & \SI{18}{ms}             \\
        \boost       & \ldots79300639239\color{red}146108\ldots & \ldots921440192\color{red}93026997\ldots & \SI{460}{\milli\second} \\
        \mathematica & \ldots793006392390\color{red}62355\ldots & \ldots9214401\color{red}6346532746\ldots & \SI{250}{\milli\second} \\
        \bottomrule
    \end{tabular}
    \caption{Numerical evaluation of the MPL in eq.~\eqref{eq:MPL_ex_2} using \ginac\ as well as the \mathematica\ implementation and the \GSL\ and \boost\ \cpp\ backends of \IterInt. Digits in red are incorrect. As the imaginary part is of the order \num{e-6}, the leading zeros have been omitted. Below the double line are the results at high precision, and the decimal expansion shown overlaps by one digit with the results above.}\label{tab:secondGres}
\end{table}

The previous example involved an MPL that required regularisation. In order to gauge the effect of the regularisation on the performance of our code, we investigate next an MPL with the same arguments, but without the need for regularisation. We consider
\beq\label{eq:MPL_ex_2}
G(0, 0, 1, 5, 3; 1 + i)\approx 0.004706\ldots+ i\,7.998\ldots \times 10^{-6}\,.
\eeq
Overall, we obtain a similar picture to the example from eq.~\eqref{eq:MPL_ex_1}. The results are shown in table~\ref{tab:secondGres}.
It can be seen that the precision of the results is higher.
This is expected, because the sum of multiple convergent integrals is generally less accurate than any single one of them.
For the imaginary part only relatively few significant digits are available. The reasons behind this are twofold.
On the one hand, its contribution to the total deviation in eq.~\eqref{eq:OptimizationTarget} is comparatively small. On the other hand, the limited floating point precision affects it much sooner.
Comparing the results at low and high precision clearly shows the expected asymptotic scaling.
For integrals like this one the other algorithms generally increase the runtime while having almost no effect on the accuracy.
This is caused by the additional overhead necessary especially for the splitting approach.
From our observations there is one notable exception to this: If integration kernels appear multiple times and are relatively expensive to compute, e.g., because arbitrary precision calculations are involved, then the tree-based approach in the \cpp\ implementations has a slightly reduced runtime with identical results, because the kernel evaluations are cached.
In this example this reduces the runtime at high precision by roughly \SI{5}{\percent}.

\begin{table}[tb]
    \centering
    \begin{tabular}{cccc}
        \toprule
        Library       & Real part                                 & Imaginary part                            & Runtime                 \\
        \midrule
        \GSL          & -0.77242\color{red}5660871518\ldots       & -1.21178\color{red}1172871826\ldots       & \SI{380}{\milli\second} \\
        \GSL*         & -0.77242\color{red}5670371146\ldots       & -1.21178\color{red}1163558710\ldots       & \SI{19}{\milli\second}  \\
        \boost        & -0.77242\color{red}6321658913\ldots       & -1.2117\color{red}79849965943\ldots       & \SI{560}{\milli\second} \\
        \boost*       & -0.77242383\color{red}1202786\ldots       & -1.21178504\color{red}4410888\ldots       & \SI{14}{\milli\second}  \\
        \mathematica  & -0.772\color{red}575263478214\ldots       & -1.21\color{red}2014140953513\ldots       & \SI{36}{s}              \\
        \mathematica* & -0.772\color{red}843167315320\ldots       & -1.21\color{red}0748406366876\ldots       & \SI{1.3}{s}             \\
        \midrule
        \midrule
        \boost        & \ldots79998952\color{red}7595750576\ldots & \ldots033593\color{red}764141561885\ldots & \SI{95}{\minute}        \\
        \boost*       & \ldots7999895200\color{red}76038830\ldots & \ldots033593841\color{red}037212356\ldots & \SI{46}{\second}        \\
        \mathematica  & \ldots799989\color{red}442070510628\ldots & \ldots033593\color{red}914143434987\ldots & \SI{32}{\minute}        \\
        \mathematica* & \ldots7999895200\color{red}16450820\ldots & \ldots0335938411\color{red}22460908\ldots                                     & \SI{30}{\minute}      \\
        \bottomrule
    \end{tabular}
    \caption{Numerical evaluation of the MPL in eq.~\eqref{eq:MPL_ex_3} using the \mathematica\ implementation and the \GSL\ and \boost\ \cpp\ backends of \IterInt. Digits in red are incorrect. The libraries denoted with * use an optimised configuration as discussed in the text. Below the double line are the results at high precision, and the decimal expansion shown overlaps by one digit with the results above.}\label{tab:longGresults}
\end{table}

Lastly, we illustrate the performance of \IterInt\ on a high-length example:
\beq\label{eq:MPL_ex_3}
G(\tfrac{3}{2}, \tfrac{3}{2}, -\tfrac{3}{7}, 1, 1, 0, 0, 0, 0, 0, 0, 0, 0, 0; 1 + 5i)\approx -0.772\ldots -i 1.211 \ldots\,.
\eeq
A comparison with \ginac\ is not possible, due to very long runtimes.
To obtain a reference result, we performed the calculation both using \cpp\ and \mathematica\ at significantly higher precision.
Table~\ref{tab:longGresults} shows runtimes and numerical results.
Here all codes yield results with comparatively low accuracy. This is caused by the large number of convergent integrals involved and the resulting increase in deviation.
It can also be seen that at high precision \mathematica\ is quite fast compared to \boost, while being slightly less accurate.
Because this is an example for a high-length iterated integral requiring regularisation, here the effects of the different algorithms can be discussed nicely.
As a first optimisation, all trivial iterated integrals produced after regularisation can be dropped. At low precision this more than halves the runtime.
With increased precision, the effect of this optimisation becomes small, because the number of steps to compute these trivially vanishing integrals remains roughly constant.
Therefore the absolute time needed to evaluate them is almost constant as well, while the time to calculate the other integrals increases much faster.

To further reduce the runtime, we can also switch to the tree-based approach. This can reduce the runtime further by up to \SI{90}{\percent} for the \cpp\ codes.
This clearly shows the amount of redundancy in the problem that is eliminated in this way. Again the accuracy of the results remains similar.
The effects on the \mathematica\ solver are slightly different. The runtime only decreases by about \SI{20}{\percent}, and the accuracy gets slightly worse.
Here probably the solver in \mathematica\ is getting overwhelmed by the complicated system with over \num{27000} functions.

Lastly, in an example like in eq.~\eqref{eq:MPL_ex_3} it is faster to use the algorithm where the path is split, so that regularisation is only applied on a short path.
The additional calculations necessary for this are comparatively small, because many of the iterated integrals necessary (or subintegrals thereof) are already being computed.
In \mathematica\ the runtime is reduced at low precision by \SI{90}{\percent}. At high precision, the effect on the runtime is miniscule.
Using \cpp\ at low precision, the runtime is reduced by roughly one third. This changes at high precision, and the runtime is reduced further by a factor of 11.
The difference in behaviour can be explained by the amount of precision-independent calculations to construct the systems of differential equations.
In table~\ref{tab:longGresults} we denote this fully optimised setup using both the two-stage, tree-based algorithms and dropping trivial integrals beforehand with a *.\footnote{The * setup used for \mathematica\ had to be slightly altered to obtain a high precision result by reducing the \texttt{WorkingPrecision} to 55 and increasing both the \texttt{AccuracyGoal} and \texttt{PrecisionGoal} to 35. This is necessary to fulfil the internal consistency checks by \mathematica.}

At high precision, we notice significant effects of the different algorithms.
Just using the tree-based approach without further optimisations reduces the runtime of the \cpp\ codes to \SI{8.75}{\minute} at a similar deviation.
The best results have been obtained by using the tree-based approach and splitting the integration region.
In this way it was possible to get a result deviating only by $10^{-24}$ in \SI{45}{\second} using the \cpp\ backend.
All tree-based approaches in \mathematica\ take about \SI{30}{\min} and yield results with a similar deviation of about $10^{-21}$.

%%%%%%%%%
\subsection{Elliptic multiple polylogarithms}\label{sec:empls}
Elliptic multiple polylogarithms (eMPLs) generalise the ordinary MPLs defined in eq.~\eqref{eq:MPL_def} to elliptic curves. The integration kernels are defined via the Kronecker-Eisenstein series:
\begin{align}
    F(z, \alpha, \tau) & \coloneqq \frac{\theta'_1(0,\tau)\theta_1(z+\alpha,\tau)}{\theta_1(z,\tau)\theta_1(\alpha,\tau)}=\sum_{n=0}^{\infty} g^{(n)}(z,\tau)\alpha^{n-1}\comma
\end{align}
where $\theta_1$ denotes a Jacobi $\theta$-function and $\theta'_1$ denotes the derivative with respect to the first argument.
eMPLs are then defined by the iterated integral~\cite{Brown:2011wfj,Broedel:2014vla,Broedel:2017kkb}
\begin{align}
    \tilde{\Gamma}\qty({}^{n_1}_{z_1} {}^{\cdots}_{\cdots} {}^{n_k}_{z_k}; z,\tau) & \coloneqq \int_{0}^{z} \dd{z'} g^{(n_1)}(z'-z_1, \tau) \tilde{\Gamma}\qty({}^{n_2}_{z_2} {}^{\cdots}_{\cdots} {}^{n_k}_{z_k}; z',\tau)\comma\label{eq:wReg}
\end{align}
where $z$, $z_i$ and $\tau$ are complex numbers with $\operatorname{Im}(\tau)>0$.
As usual, the empty integral is set to unity. In the notations of section~\ref{sec:iterated}, we have
\beq
\tilde{\Gamma}\qty({}^{n_1}_{z_1} {}^{\cdots}_{\cdots} {}^{n_k}_{z_k}; z,\tau) = I(\omega_{n_k,z_k},\ldots,\omega_{n_1,z_1};z)\,,\textrm{~~~with~~~} \omega_{n_i,z_i} = g^{(n_i)}(z'-z_i,\tau)\dd{z'}\fullstop
\eeq
To numerically evaluate the integration kernels $g^{(n)}(z,\tau)$, we use the series expansions
\begin{equation}\begin{split}
    g^{(0)}(z,\tau)     & =1\comma                                                                                                                        \\
    g^{(1)}(z,\tau)     & =\pi \cot(\pi z) + 4 \pi \sum_{m=1}^{\infty}\sin(2 \pi m z) \sum_{n=1}^{\infty} q^{m n}\comma                                   \\
    g^{(2 k)}(z,\tau)   & =-2\qty[\zeta(2k)+\frac{{(2\pi i)}^{2k}}{(2k-1)!}\sum_{m=1}^{\infty}\cos(2 \pi m z) \sum_{n=1}^{\infty} n^{2k-1} q^{m n}]\comma \\
    g^{(2 k+1)}(z,\tau) & =-2i\frac{{(2\pi i)}^{2k+1}}{(2k)!}\sum_{m=1}^{\infty}\sin(2 \pi m z) \sum_{n=1}^{\infty} n^{2k} q^{m n}\fullstop
\end{split}\end{equation}
Here, $\zeta$ denotes the Riemann zeta function and $q=e^{2\pi i \tau}$.
Note that the infinite sum over $n$ can be expressed as a rational function in $q^m$, because
\begin{align}
    {\qty(x\dv{x})}^k \frac{x}{1-x} & =\sum_{n=1}^{\infty} n^k x^n\,.
\end{align}
Since the function $g^{(n)}(z,\tau)$ can have singularities for $z\in \Z+\tau\Z$, these series expansions do not converge globally.
To obtain numerical results for arbitrary values of $z$ and $\tau$, the quasi-periodicity of $F$ in $z$ is used,
\begin{align}
    F(z+n+m\tau,\alpha,\tau) & =e^{-2\pi i \alpha m}F(z, \alpha,\tau)\comma\qquad n,\,m\in\Z\,.
\end{align}
This allows us to translate $z$ back into the region of convergence.
For all of the following calculations the series expansions of $g^{(n)}(z,\tau)$ are truncated at $m=10$ for low precision and $m=20$ for high precision calculations, giving roughly 28 and 56 digits of precision respectively for $\Im\tau\approx 1$.

There are two different ways in the literature to regulate eMPLs.
The first is simply to introduce a tangential base-point for the integrals in $z$, as described in section~\ref{sec:iterated}. The second one introduces the tangential base-point at 1 after changing variables to $w=e^{2\pi i z}$. This leads to~\cite{Broedel:2014vla,Broedel:2017jdo}
\begin{align}\label{eq:eMPL_w_reg}
    \tilde{\Gamma}\qty({}^1_0; z, \tau) & = \log(1-e^{2\pi i z})-2\pi i z+\int_0^z \dd{z'}\qty[g^{(1)}(z',\tau)-\frac{2\pi i}{e^{2\pi i z'}-1}]\,.
\end{align}
The two regularisations differ, and lead to different numerical values for integrals that require regularisation.
For consistency, we will follow the regularisation convention of regulating in $z$, i.e., we follow closely the prescription of section~\ref{sec:iterated}. We show how to related results obtained with these two regularisations in appendix~\ref{app:regularisation}.

Currently, there are two public libraries that allow one to evaluate eMPLs numerically. First, ref.~\cite{Walden:2020odh} allows for the computation of large classes iterated integrals via series expansions, and it includes an implementation of eMPLs into \ginac. This code, however, is restricted to values of the arguments that lie inside the range of convergence of the series expansions.
Recently, another method, including an implementation into \ginac, was introduced, based on obtaining a series expansion of eMPLs in $q=e^{2\pi i \tau}$ whose coefficients are ordinary MPLs~\cite{Duhr:2026ell}.
In the remainder of this section, we present a comparison of \IterInt\ and the \ginac\ implementation of eMPLs from refs.~\cite{Walden:2020odh,Duhr:2026ell}. We only compare our \cpp\ implementations with the \ginac\ code. We use the same precision goals as in the previous section.
We note that, since this code uses the regularisation according to eq.~\eqref{eq:wReg}, the numerical values will differ if the integrals need to be regularised. We can relate them using the results from appendix~\ref{app:regularisation}. In addition to the explicit examples discussed in the remainder of this section, to validate our code we have compared the output of \IterInt\ for a few hundred eMPLs evaluated at random real arguments against \ginac, and we always find very good agreement.

\begin{table}[tb]
    \centering
    \begin{tabular}{c c c}
        \toprule
        Library     & Value                                     & Runtime                 \\
        \midrule
        \GSL        & -0.011625009667135\color{red}\ldots       & \SI{79}{\micro\second}  \\
        \boost      & -0.01162500966\color{red}3896\ldots       & \SI{84}{\micro\second}  \\
        \ginac~\cite{Walden:2020odh} & -0.011625009667135\ldots                  & \SI{125}{ms}            \\
        \ginac~\cite{Duhr:2026ell} & -0.011625009667135\ldots                  & \SI{630}{\micro\second} \\
        \midrule
        \midrule
        \boost      & \ldots571402160388954611\color{red}\ldots & \SI{580}{\milli\second} \\
        \ginac~\cite{Walden:2020odh} & \ldots571402160388954611\ldots            & \SI{144}{ms}            \\
        \ginac~\cite{Duhr:2026ell} & \ldots571402160388954611\ldots            & \SI{4}{ms}              \\
        \bottomrule
    \end{tabular}
    \caption{Numerical evaluation of the eMPL in eq.~\eqref{eq:EMPL_ex_1} using \ginac\ as well as the \GSL\ and \boost\ \cpp\ backends of \IterInt. Digits in red are incorrect.}\label{tab:EMPLfirstRes}
\end{table}

As a first example we consider example 3 from ref.~\cite{Walden:2020odh} and calculate $\tilde{\Gamma}\qty({}^0_0 {}^1_{z_1}; z, \tau)$ with $z_1=\inv{3}$, $z=\inv{10}$, and $\tau=i$. The numerical value is
\beq\label{eq:EMPL_ex_1}
\tilde{\Gamma}\qty({}^0_0 {}^1_{z_1}; z, \tau) = -0.011625\ldots\fullstop
\eeq
\IterInt\ obtains a result in less than \SI{100}{\micro\second}, and it deviates by \num{3e-12} (\boost) or \num{2e-16} (GSL) from the benchmark value obtained by multiple calculations with very high precision. The \ginac\ implementation based on ref.~\cite{Walden:2020odh} requires about {1\,500}~times more time at a similar accuracy.\footnote{We have noticed that, for a reason that we ignore, in a given \ginac\ program the first integration always takes significantly longer. To get consistent runtimes, we always perform a dummy calculation first, and we only report the \ginac\ runtimes for the subsequent runs.}
After increasing the precision, \ginac\ is faster by a factor of about four with a more accurate result. Our code still gives a result with a deviation below $10^{-33}$.
Comparing to the \ginac\ implementation based on ref.~\cite{Duhr:2026ell}, at low precision the runtime of our approach is smaller by more than a factor of 7. At high precision this code is roughly 145 times faster than ours.
The results are summarised in table~\ref{tab:EMPLfirstRes}.

\begin{table}[tb]
    \centering
    \begin{tabular}{c c c}
        \toprule
        Library     & Value                                     & Runtime                 \\
        \midrule
        \GSL        & -5.5117484\color{red}90990788\ldots       & \SI{200}{\micro\second} \\
        \boost      & -5.\color{red}447618524308661\ldots       & \SI{240}{\micro\second} \\
        \boost*     & -5.51174846551103\color{red}3\ldots       & \SI{122}{\milli\second} \\
        \ginac~\cite{Walden:2020odh} & -5.51174846\color{red}4595990\ldots        & \SI{63}{\second}        \\
        \ginac~\cite{Duhr:2026ell} & -5.511748465\color{red}380212\ldots       & \SI{363}{\milli\second} \\
        \midrule
        \midrule
        \boost      & \ldots02713617874\color{red}6934756\ldots & \SI{4.3}{\second}       \\
        \ginac~\cite{Walden:2020odh} & -5.51174846\color{red}4595982\ldots       & \SI{5.5}{\hour}         \\
        \ginac~\cite{Duhr:2026ell} & \ldots027136178744\color{red}517954\ldots & \SI{1.5}{\second}       \\
        \bottomrule
    \end{tabular}
    \caption{Numerical evaluation of the eMPL in eq.~\eqref{eq:EMPL_ex_2} using \ginac\ as well as the \GSL\ and \boost\ \cpp\ backends of \IterInt. Digits in red are incorrect. We rescaled all results by $10^{10}$. At high precision, the \ginac\ implementation of ref.~\cite{Walden:2020odh} did not yield results of higher precision than at low precision. The row marked with * uses arbitrary precision calculations to obtain a sensible amount of precision.}\label{tab:EMPLSecondRes}
\end{table}

Our code is especially efficient to evaluate iterated integrals of high length.
As an example, we evaluate
\beq\label{eq:EMPL_ex_2}
\tilde{\Gamma}\qty({}^1_{z_1} {}^2_{z_2} {}^3_{z_3} {}^4_{z_4} {}^3_{z_3} {}^2_{z_2} {}^1_{z_1}; z, \tau)\approx -5.512\ldots \times 10^{-10}\,,
\eeq
where $z_i=\inv{i+2}$, $z=\inv{10}$, and $\tau=i$.
Let us start again by discussing the low precision results. We note that low precision calculations have only very limited usability, because an absolute deviation of \num{1e-12} would correspond to two correct digits.\footnote{Requiring a corresponding relative precision is not really possible at double precision, as problems due to floating point arithmetic start to appear quickly.} \GSL\ produces 8 correct digits in \SI{200}{\micro\second}.
When switching to arbitrary precision arithmetic and choosing $\varepsilon_\text{abs}=10^{-22}$, we obtain 14 correct digits in \SI{122}{\milli\second}.
The \ginac\ implementation based on ref.~\cite{Walden:2020odh} obtains 10 significant digits immediately, but in a runtime of slightly more than one minute, while the implementation based on ref.~\cite{Duhr:2026ell} takes more than {1\,800} times longer than \GSL.
If we pass to higher accuracy, the runtime increases to \SI{5.5}{\hour} for ref.~\cite{Walden:2020odh}, while for ref.~\cite{Duhr:2026ell} it is roughly three times shorter than for \IterInt.
\IterInt\ produces results deviating by less than \num{2e-36} in \SI{4.3}{\second}.
These results can be found in table~\ref{tab:EMPLSecondRes}.

\begin{table}[tb]
    \centering
    \begin{tabular}{c c c c}
        \toprule
        Library     & Real part                                 & Imaginary part                            & Runtime                \\
        \midrule
        \GSL        & -4.93407984\color{red}3678764\ldots       & -9.97494627\color{red}7007349\ldots       & \SI{21}{\milli\second} \\
        \boost      & -4.9340798\color{red}22720885\ldots       & -9.9749462\color{red}50109483\ldots       & \SI{16}{\milli\second} \\
        \ginac~\cite{Duhr:2026ell} & -4.934079842911309\ldots                  & -9.974946276588365\ldots                  & \SI{13.1}{\minute}     \\
        \midrule
        \midrule
        \boost      & \ldots948960857090\color{red}836177\ldots & \ldots5485351648329\color{red}24647\ldots & \SI{24.4}{\minute}     \\
        \ginac~\cite{Duhr:2026ell} & \ldots948960857090652\color{red}411\ldots & \ldots548535164832980\color{red}875\ldots & \SI{36.4}{\minute}     \\
        \bottomrule
    \end{tabular}
    \caption{Numerical evaluation of the eMPL in eq.~\eqref{eq:EMPL_ex_2} using the \ginac\ implementation of ref.~\cite{Duhr:2026ell} as well as the \GSL\ and \boost\ \cpp\ backends of \IterInt. Digits in red are incorrect. For consistency within this table we chose the regularisation used by \ginac\ for \IterInt\ as well. The numerical values have been rescaled by $10^{-17}$.}\label{tab:EMPLThirdRes}
\end{table}

Finally, let us discuss an eMPL where the approach via series expansion is not applicable.
We consider the function:
\begin{align}
    \tilde{\Gamma}\qty({}^4_{w_1} {}^3_{w_2} {}^1_0 {}^1_0; w,\tau) & \approx -(4.86\ldots+i\,9.934 \ldots) \times 10^{17}\,,
\end{align}
with $w_1=\tfrac{73}{3}$, $w_2=\tfrac{152}{13}+3 i=3\tau + 2$,  $w=\tfrac{13}{7}+\tfrac{195}{11}i$ and $\tau=\tfrac{42}{13}+i$. Thus, three integration kernels are involved in the regularisation, and we use the tree-based approach.
The \ginac\ code of ref.~\cite{Duhr:2026ell} converts the single integral into a large number of subintegrals, resulting in a very significant memory consumption of about \SI{7}{\giga\byte}.
Because the resulting value is quite large, it is necessary to change to a relative requested precision of \num{1e-12} and \num{1e-30} respectively instead.
\GSL\ requires a runtime of \SI{21}{\milli\second} compared to the \SI{13.1}{\minute} for ref.~\cite{Duhr:2026ell}.
The result obtained by \GSL\ has a relative deviation of roughly \num{1e-10}, while the code of ref.~\cite{Duhr:2026ell} reaches \num{1e-16}.
As usual, the \boost\ backend is slightly less accurate at \num{1e-9} relative deviation, but significantly faster.
At high precision our approach takes \SI{24.4}{\minute} with a relative deviation of \num{2e-28} compared to the \SI{36.4}{\minute} of the code of ref.~\cite{Duhr:2026ell} at a relative deviation of less than \num{1e-31}.
These results can also be found in table~\ref{tab:EMPLThirdRes}.

%%%%%%%%%%%%%%%%%%%%%%%%%%

\section{Numerical evaluation of banana integrals with up to four loops}\label{sec:examples}

In this section we want to showcase the usage of \IterInt\ to evaluate the iterated integrals that arise from differential equations in canonical form satisfied by certain multi-loop Feynman integrals. We focus on banana integrals with massive propagators with up to four loops (and we exclude the one-loop case, which can be expressed in terms of MPLs) in $D=2-2\epsilon$ dimensions, which are related to the results in $D=4-2\epsilon$ via dimensional shift relations~\cite{Tarasov:1996br,Lee:2009dh}.

\subsection{Generalities}

The $l$-loop banana integral family in $D=2-2\epsilon$ dimensions is defined by\footnote{In order to simplify the discussion, we exclude numerator factors from the definition. This is not a restriction, because we can always pick a basis of master integrals without numerators.}
\begin{align}
    I_{\nu_1\dots\nu_{l+1}}(D,p^2,m_1^2,\dots,m_{l+1}^2) & =e^{l\gamma\epsilon}\int\frac{\dd[D]{k_1}\cdots\dd[D]{k_l}}{{(i\pi^\frac{D}{2})}^l}\inv{D_1^{\nu_1}\cdots D_{l+1}^{\nu_{l+1}}}\comma\label{eq:defLLoopBanana}
\end{align}
where $\gamma=-\Gamma'(1)$ denotes the Euler-Mascheroni constant. The propagators are defined by
\begin{equation}\begin{split}
        D_i     & =k_i^2-m_i^2\quad\text{for}\quad 1\leq i\leq l\comma \\
        D_{l+1} & ={\qty(p-\sum_{i=1}^l k_i)}^2-m_{l+1}^2\fullstop
    \end{split}\end{equation}
Using integration-by-parts (IBP) identities~\cite{Tkachov:1981wb,Chetyrkin:1981qh}, every integral from this family can be expressed in terms of a finite number $N$ of basis integrals, commonly referred to as \emph{master integrals} in the literature. The number $N$ of master integrals depends on the number of loops and on whether some of the masses are equal (see, e.g., ref.~\cite{Duhr:2026elp} for the number of master integrals up to four loops for any mass configuration). If we package the $N$ master integrals into a vector $\mathbf{I}$, then this vector satisfies a system of first-order linear differential equations~\cite{Kotikov:1990kg,Kotikov:1991hm,Kotikov:1991pm,Gehrmann:1999as,Henn:2013pwa},
\beq\label{eq:DEQ_FI}
\mathrm{d}\mathbf{I}(\mathbf{x},\epsilon) = \mathbf{A}(\mathbf{x},\epsilon)\,\mathbf{I}(\mathbf{x},\epsilon)\,,
\eeq
where $\mathbf{x}$ is the vector of independent ratios of the form $\tfrac{p^2}{m_i^2}$, and $\mathbf{A}(\mathbf{x},\epsilon)$ is a matrix of rational functions in $\epsilon$ and of rational one-forms in $\mathbf{x}$.
The differential equation in eq.~\eqref{eq:DEQ_FI} may be difficult to solve. However, we are only interested in the solution as a Laurent expansion in the dimensional regulator $\epsilon$ around zero. To find such a solution, it turns out to be convenient to change basis according to $\mathbf{J}(\mathbf{x},\epsilon) = \mathbf{R}(\mathbf{x},\epsilon)\mathbf{I}(\mathbf{x},\epsilon)$, such that the new vector of master integrals satisfies a differential equation in \emph{canonical form}, where in particular the dimensional regulator $\epsilon$ only enters in a factorised manner:
\beq\label{eq:canonical_DEQ}
\mathrm{d}\mathbf{J}(\mathbf{x},\epsilon) = \epsilon\,\mathbf{\widetilde{A}}(\mathbf{x})\,\mathbf{J}(\mathbf{x},\epsilon)\,.
\eeq
This special form of the differential equation was first pointed out in ref.~\cite{Henn:2013pwa} in the context of Feynman integrals that evaluate to MPLs, in which case the matrix $\mathbf{\widetilde{A}}(\mathbf{x})$ is a matrix of dlog-forms. Various approaches to obtain differential equations in canonical form for functions beyond MPLs have been recently proposed~\cite{Pogel:2022ken,Pogel:2022yat,Pogel:2022vat,Gorges:2023zgv,Duhr:2025lbz,Maggio:2025jel,e-collaboration:2025frv,Bree:2025tug,Chen:2025hzq,Yang:2025ofz,Forner:2026vby}, and it is expected that $\mathbf{\widetilde{A}}(\mathbf{x})$ only has simple poles (at least locally)~\cite{Duhr:2025lbz}. The general solution of eq.~\eqref{eq:canonical_DEQ} can then be cast in the form of a path-ordered exponential,
\begin{align}
    \mathbf{J}(\mathbf{x},\epsilon) & =\operatorname{\mathbb{P}exp}\qty(\epsilon\int_\gamma\mathbf{\widetilde{A}}(\mathbf{x}))\mathbf{J}_0(\epsilon)\,,
\end{align}
where $\gamma$ is a path from a boundary point $\mathbf{x}_0$ to a generic point $\mathbf{x}$, and $\mathbf{J}_0(\epsilon)$ is the (regulated) value of $\mathbf{J}(\mathbf{x},\epsilon)$ at $\mathbf{x}=\mathbf{x}_0$. We can easily expand the path-ordered exponential into a series in $\epsilon$, and the coefficients of the Laurent expansion are iterated integrals over the letters that appear in the matrix $\mathbf{\widetilde{A}}$. The resulting iterated integrals will always be homotopy-invariant. Hence, we see that solving systems of differential equations in canonical form satisfied by Feynman integrals naturally leads to special functions that can be evaluated using \IterInt.
The aim of this section is to illustrate this on the example of banana integrals with two, three or four loops.

Differential equations in canonical form for banana integrals with restricted assignments of the masses have been obtained in refs.~\cite{Pogel:2022ken,Pogel:2022yat,Pogel:2022vat,Maggio:2025jel,Duhr:2025kkq,Pogel:2025bca,Duhr:2025ouy}.
It is well known that beyond one loop, banana integrals are associated with Calabi-Yau (CY) geometries. More precisely, at $l$ loops the maximal cut of the master integral $I_{1\cdots1}$, which is a solution to the homogeneous equation associated with the system in eq.~\eqref{eq:DEQ_FI}~\cite{Primo:2016ebd,Bosma:2017hrk,Frellesvig:2017aai}, computes a period of a CY $(l-1)$-fold~\cite{Sabry,Caffo:1998du,Laporta:2004rb,Laporta:2008sx,Muller-Stach:2011qkg,Muller-Stach:2012tgj,ell2,Bloch:2014qca,Bloch:2016izu,Klemm:2019dbm,Bonisch:2020qmm}. As a consequence, the solutions to the differential equations will involve integration kernels that contain the (quasi-)periods of the CY variety, and these iterated integrals can typically not be evaluated using standard libraries for special functions. As we will see, once the periods of the CY variety are known (and can be evaluated numerically), we can easily use \IterInt\ to evaluate the relevant iterated integrals.

\subsection{The two-loop sunrise integral}
Let us start by discussing the two-loop banana integral, also known as the sunrise integral. This integral was expressed in terms of the eMPLs defined in section~\ref{sec:empls} in refs.~\cite{Broedel:2017siw,Campert:2020yur}, and in the equal-mass case in terms of iterated integrals of modular forms~\cite{ManinModular,Brown:2014pnb} in refs.~\cite{Adams:2017ejb,Broedel:2018iwv}.

\paragraph{The unequal-mass sunrise integral.}
In the case of three distinct non-zero propagator masses, there are seven master integrals, three of which can be written as products of one-loop tadpole integrals. A system of differential equations in canonical form was obtained in ref.~\cite{Bogner:2019lfa}. We will be brief and not discuss the differential equation and its initial condition, and we refer to ref.~\cite{Bogner:2019lfa} for the details. Rather, we only describe the letters that enter the differential equations and how to implement them into \IterInt.

It is by now well known that the two-loop sunrise integral is associated to a family of elliptic curves. It is then useful to change to variables that are naturally associated with an elliptic geometry. In particular, in ref.~\cite{Bogner:2019lfa} it was shown that the three independent kinematics variables $\mathbf{x}$ can naturally encoded into the modular parameter $\tau$ (with Im $\tau>0$) and three special points $z_i\in \mathbb{C}/\Lambda$, with $1\le i\le 3$ and $z_1+z_2+z_3=1$. $\mathbb{C}/\Lambda$ is the elliptic curve defined as the lattice $\Lambda=\mathbb{Z}+\mathbb{Z}\tau$. For the precise form of the change of variables, we refer to ref.~\cite{Bogner:2019lfa}. With these variables, the letters can be cast in the form
\begin{align}\label{eq:omega_letters_sunrise}
    \omega_k(z_i,\tau) & = {(2\pi)}^{2-k}\left[g^{(k-1)}(z_i,\tau)\dd{z_i}+(k-1)g^{(k)}(z,\tau)\frac{\dd{\tau}}{2\pi i}\right]\,,
\end{align}
where the functions $g^{(k)}(z,\tau)$ have been defined in section~\ref{sec:empls} and we use the convention $g^{(-1)}(z,\tau)\equiv 0$.
In addition, there are kernels that only depend on $\tau$,
\begin{equation}\begin{split}
\label{eq:sunrise_diff_mass_eisenstein}
        %        b_{2}(\tau)  & = \qty[e_2 (\tau)-2e_2 (2\tau)]\,\dd{\tau}\comma            \\
        \eta_2(\tau) & = \qty[e_2 (\tau)-2e_2 (2\tau)]\,\frac{\dd{\tau}}{2\pi i}\comma \\
        \eta_4(\tau) & = \inv{{(2\pi i)}^2}e_4(\tau)\frac{\dd{\tau}}{2\pi i}\comma
    \end{split}\end{equation}
where we defined the Eisenstein series of weight $k$,
\begin{equation}
    e_k(\tau)     = \sideset{}{'}{\sum_{a, b}}\inv{{(a+b\tau)}^k}\comma
\end{equation}
with $\sum'_{a, b}$ we denote the sum over $\Z^2\setminus\{(0,0)\}$, where the sum over the first component is performed first.

Let us now discuss how we can evaluate the iterated integrals that arise from the differential equations satisfied by the master integrals for the sunrise family using \IterInt. We only need to implement routines to evaluate the integration kernels in eqs.~\eqref{eq:omega_letters_sunrise} and~\eqref{eq:sunrise_diff_mass_eisenstein}. The evaluation of the functions $g^{(n)}$ was already discussed in section~\ref{sec:empls}.
To evaluate the integration kernels in eq.~\eqref{eq:sunrise_diff_mass_eisenstein}, we use the well-known expansion of Eisenstein series $e_k$ as a series in $q=e^{2\pi i \tau}$,
\begin{align}
    e_k(\tau) & =2\zeta(k)+2\frac{{\qty(2\pi i)}^k}{(k-1)!}\sum_{m=1}^\infty\sigma_{k-1}(m)q^m\,,\quad\text{with}\quad \sigma_l(m)=\sum_{d|m}d^l\comma
\end{align}
and we truncate after a fixed number of terms.
To ensure an adequate number of terms has been used, we compare the numerical result with an expansion to significantly higher order at $\tau=0.4 i$, which roughly corresponds to the smallest imaginary part appearing in our calculations.

\begin{figure}[t]
    \centering
    \includegraphics[width=0.49\textwidth]{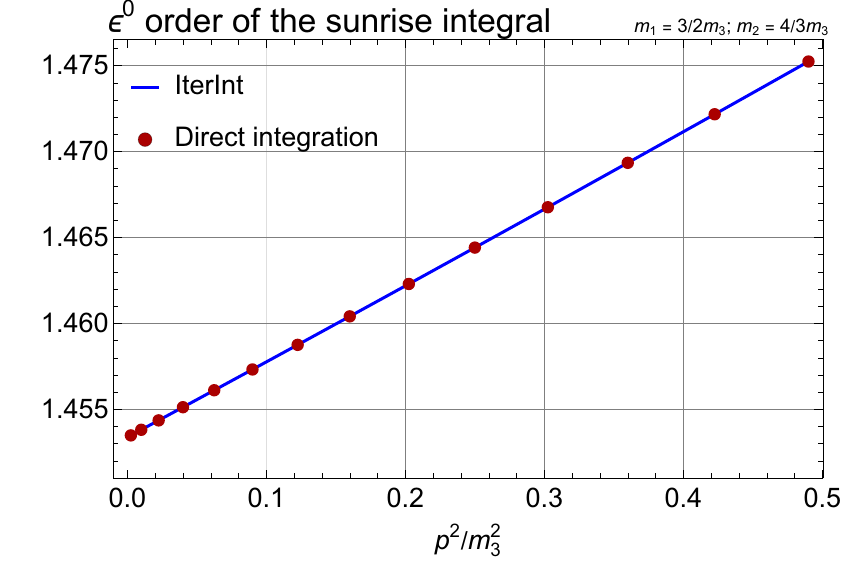}
    \includegraphics[width=0.49\textwidth]{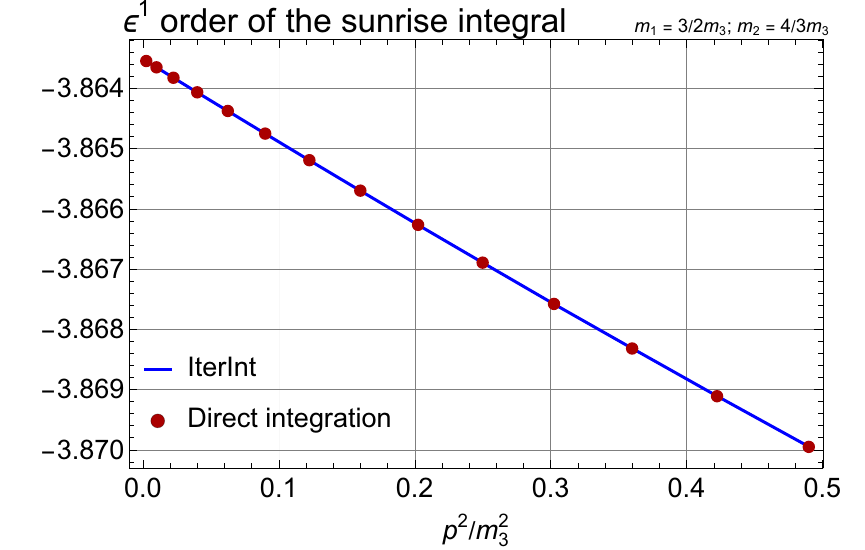}
    \includegraphics[width=0.49\textwidth]{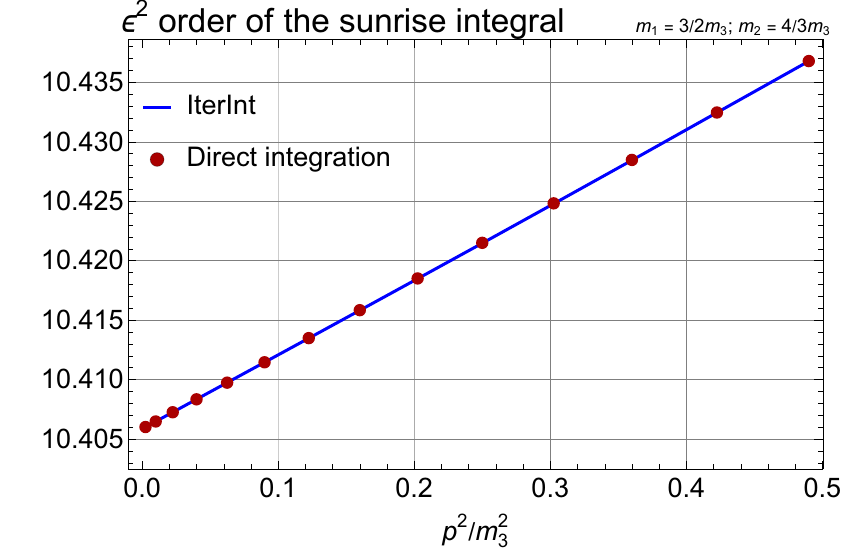}
    \caption{Numerical evaluation of the unequal-mass two-loop sunrise integral $I_{111}$ (blue line). The red dots correspond to an independent evaluation of the integral by numerically integrating the Feynman parameter representation. Note that the range on the $x$-axis corresponds to the region below threshold, where the integral is real.
    }\label{fig:sunriseComp}
\end{figure}

After these expressions have been implemented (either in \mathematica\ or in \cpp), we can immediately evaluate all the relevant iterated integrals numerically.
We have solved the differential equation from ref.~\cite{Bogner:2019lfa} to obtain an expression for the first few orders in the Laurent expansion in $\epsilon$ of the master integrals for the two-loop sunrise family in terms of iterated integrals over the aforementioned kernels.
Our results for the first three orders of the master integral $I_{111}$ are presented in figure~\ref{fig:sunriseComp}.
We present results for the one-dimensional slice where $m_1=\tfrac{3}{2}\,m_3$ and $m_2=\tfrac{4}{3}\,m_3$, and we vary the ratio $p^2/m_3^2$. 
We also include the results obtained by a direct numerical integration of the Feynman parameter representation, and we find very good agreement.

\begin{figure}[t]
    \centering
    \includegraphics[width=0.49\textwidth]{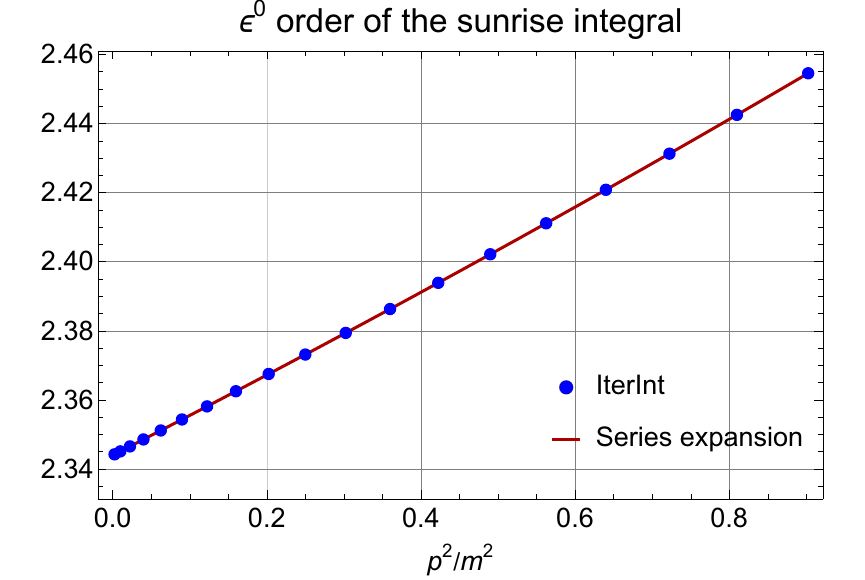}
    \includegraphics[width=0.49\textwidth]{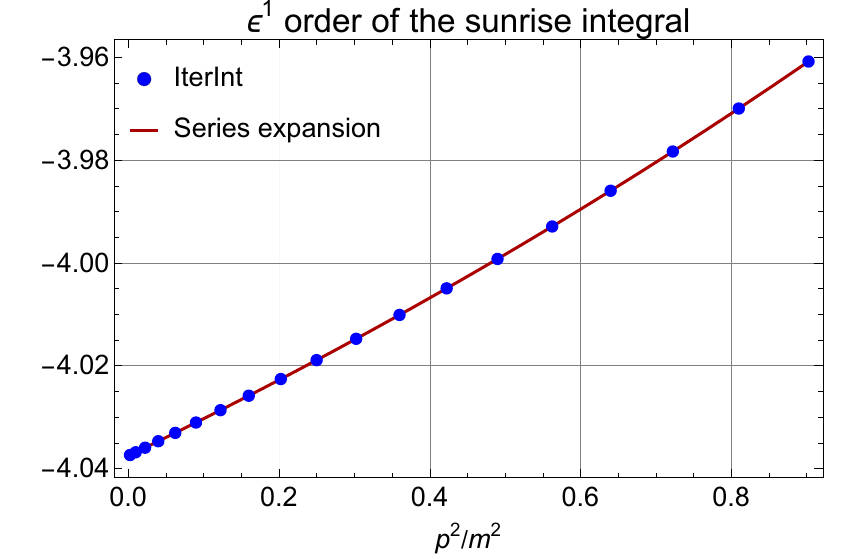}
    \includegraphics[width=0.49\textwidth]{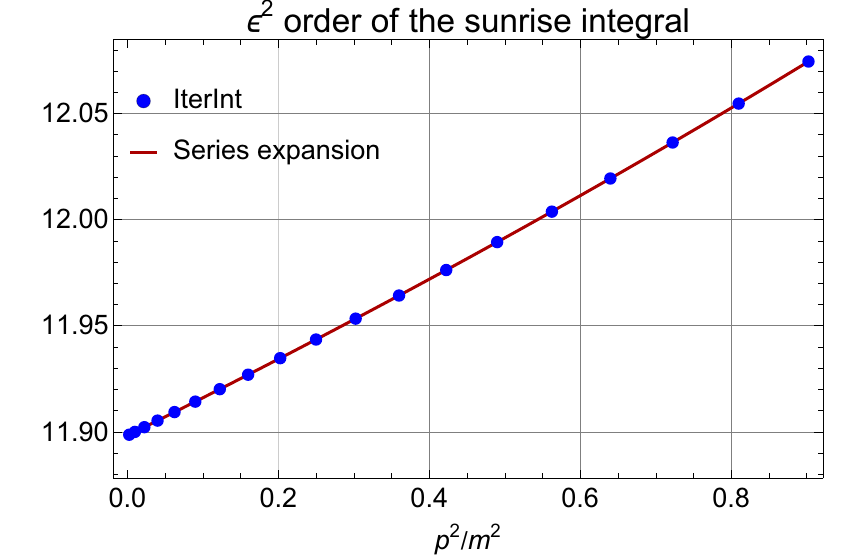}
    \caption{Numerical evaluation using \IterInt\ of the master integral $I_{111}$ of the equal-mass sunrise family (blue dots). The red line correspond to an independent evaluation of the integral using a series representation of the result. Note that the range on the $x$-axis corresponds to the region below threshold, where the integral is real. 
    }\label{fig:sunriseCompEqMass}
\end{figure}

\paragraph{The equal-mass sunrise integral.}
If all three masses are equal, there are only three master integrals (one of them is a product of one-loop tadpole integrals), and a system of differential equations in canonical form was obtained in ref.~\cite{Adams:2018yfj}.  The letters can be expressed in terms of Eisenstein series for the congruence subgroup $\Gamma_1(6)$~\cite{Adams:2017ejb}. For example, the following functions appear in the definition of the integration kernels,
\begin{equation}
    \begin{aligned}
        \Psi_1&=\frac{2\pi}{\sqrt{3}}\frac{{\eta(\tfrac{\tau}{2})}^3{\eta(2\tau)}^3 {\eta(3\tau)}}{{\eta(\tau)}^3{\eta(\tfrac{3}{2}\tau)}\eta(6\tau)}\comma\\
%        x&=-9{\qty(\frac{\eta(\tau)\eta(\tfrac{3}{2}\tau)\eta(6\tau)}{\eta(\tfrac{1}{2}\tau)\eta(2\tau)\eta(3\tau)})}^4\comma\\
%        W&=-\frac{12\pi i}{m^2 x(x-1)(x-9)}\comma\\
%        f_1&=\frac{x+3}{\sqrt{24}}\frac{\Psi_1}{\pi}\comma\\
%        f_2&=\frac{\Psi_1^2}{i\pi W}\frac{3x^2-10x-9}{2 m^2 x(x-1)(x-9)}\comma\\
        f_3&=3\sqrt{3}\frac{{\eta(\tau)}^{11}{\eta(3\tau)}^7}{{\eta(\tfrac{1}{2}\tau)}^5{\eta(2\tau)}^5\eta(\tfrac{3}{2}\tau)\eta(6\tau)}\comma
    \end{aligned}
\end{equation}
where $\eta(\tau)$ denotes the Dedekind eta function,
\begin{align}\label{eq:Dedekind}
    \eta(\tau) & = q^{1/24}\prod_{n=1}^{\infty}(1-q^n)\quad\text{with}\quad q=e^{2\pi i \tau}\fullstop
\end{align}
We could again proceed as in the case of different masses, and implement a truncated $q$-expansion for the integration. In order to showcase the flexibility of the code, however, we proceed in a different fashion. Indeed, we may exploit the fact that \mathematica\ has an implementation of the Dedekind eta function. Hence, if we work with the \mathematica\ implementation of \IterInt, we can easily implement all the kernels. They in turn were then evaluated using the built-in routines in \mathematica\ ensuring sufficient numerical precision. We compared to the truncated series representation obtained starting from the results from ref.~\cite{Adams:2017ejb}, and we found very good agreement. The result is shown in figure~\ref{fig:sunriseCompEqMass}.

\subsection{The three-loop equal-mass banana integral}
As a second example, we consider the three-loop equal-mass banana integral. This family has four master integrals, one of which is a product of one-loop tadpole integrals. The geometry associated to the three-loop banana family (also in the unequal-mass case) is a family of K3 surfaces. In the special case where all the masses are equal, the periods of this family of K3 surfaces can be expressed as products of the same family of elliptic curves as for the two-loop sunrise integral~\cite{Primo:2017ipr,Bloch:2016izu}. In refs.~\cite{Bloch:2016izu,Broedel:2019kmn,Broedel:2021zij} it was shown that this integral can be expressed via iterated integrals of meromorphic modular forms.

A system of differential equations in canonical form satisfied by the master integrals was obtained in ref.~\cite{Pogel:2022yat}. If we change variables from $x=\tfrac{p^2}{m^2}$ to the modular parameter $\tau$, then the letters can be expressed in terms of meromorphic modular forms for the congruence subgroup $\Gamma_1(6)$, as well as one letter that is itself an integral of a magnetic modular form~\cite{magnetic1,magnetic3,Bonisch:2024nru}.
We have implemented these kernels into \IterInt,
by again expressing the modular forms in terms of $\eta$-quotients. The resulting expressions were then evaluated using built-in functionality of \mathematica.

\begin{figure}[t]
    \centering
    \includegraphics[width=0.49\textwidth]{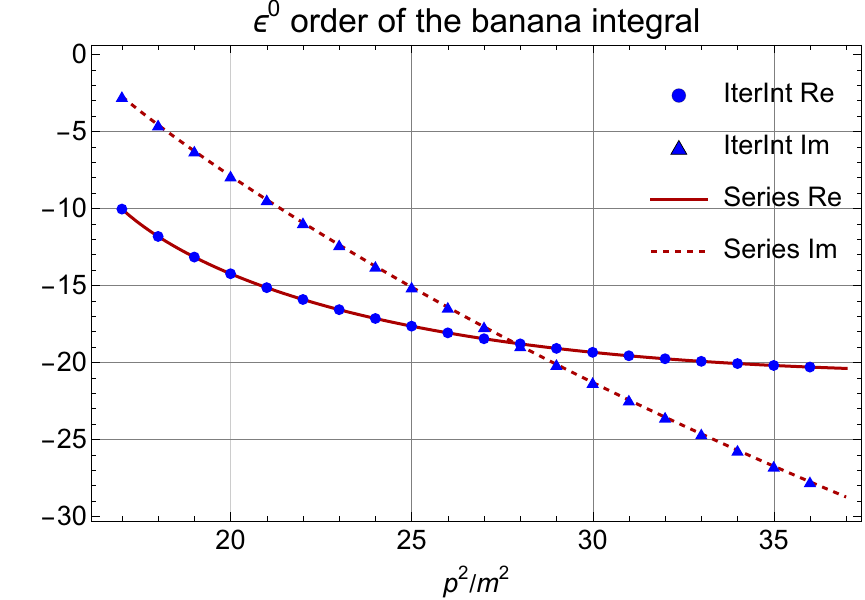}
    \includegraphics[width=0.49\textwidth]{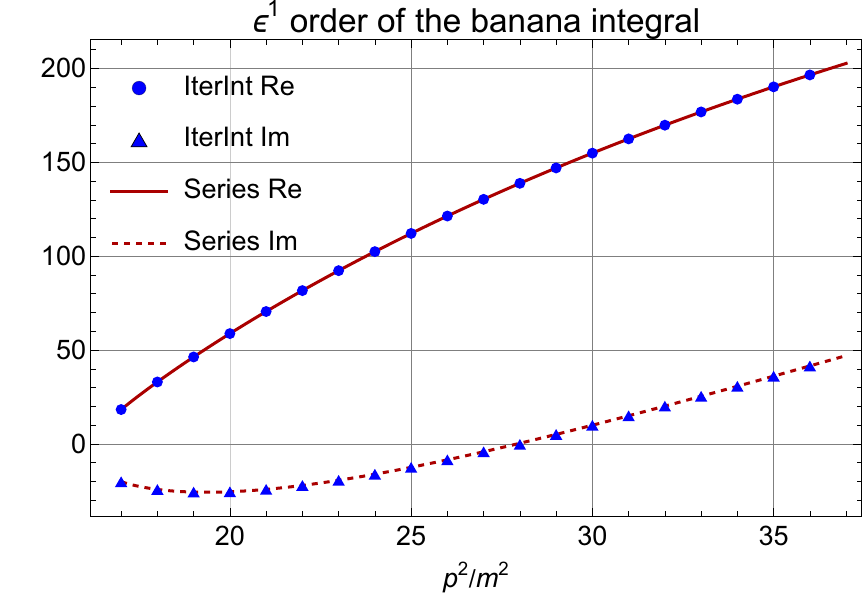}
    \caption{Numerical evaluation using \IterInt\ of the master integral $I_{1111}$ of the equal-mass banana family (blue dots and triangles). The red lines correspond to an independent evaluation of the integral using a truncated series expansion. Note that the range on the $x$-axis corresponds to the region above threshold, and both the real (solid lines) and imaginary (dashed line) parts are shown. 
}\label{fig:bananaComp}
\end{figure}

With the kernels implemented, we can easily evaluate all the iterated integrals that arise from solving the system of canonical differential equations for the equal-mass three-loop banana integrals. The results for the first two orders of the master integral $I_{1111}$ is shown in figure~\ref{fig:bananaComp}, and we observe again very good agreement compared to the results obtained with the truncated series expansion provided with ref.~\cite{Pogel:2022yat}.

Let us conclude the discussion of this example by making a comment. We already mentioned that one of the letters is itself expressed as an iterated integral~\cite{Pogel:2022yat}. This is in fact not a coincidence, but rather this is a general feature of differential equations in canonical forms, cf.~refs.~\cite{Duhr:2025lbz,Gorges:2023zgv,e-collaboration:2025frv,Bree:2025tug}. While in some instances these iterated integrals may be related to other classes of functions (cf., e.g., refs.~\cite{Duhr:2024uid,Duhr:2025xyy}), in general it is expected that these iterated integrals define new classes of special functions. We now briefly comment on how to handle such objects within \IterInt. If this kernel admits a convergent series representation, we may just implement a truncated series in the same way as for the other kernels involving, e.g., the periods of the underlying geometry. This is the approach we followed here, but it is only applicable is a converging series representation is available. An alternative approach, which does not rely on the existence of a series representation, is to expand the iterated integrals using the shuffle algebra in a way that only integration kernels without additional integrations remain. This can be achieved using relations like
(we suppress the integration boundaries for readability):
\begin{align}
    \nonumber    I(f_1,\dots,f_k,I(g_1,\dots,g_l),f_{k+1},\dots, f_n) & \,=I(f_1,\dots,f_k, I(g_1,\dots,g_l)\cdot I(f_{k+1},\dots,f_n))\,, \\
    I(f_1,\dots, f_k, I(f_{k+1},\dots,f_n))                           & \,=I(f_1,\dots,f_k,1,f_{k+1},\dots,f_n)\,.
\end{align}
While this strategy is always applicable, the computational effort may increase significantly, because the number of iterated integrals that need to be evaluated may substantially increase after expanding out all shuffle products.

\subsection{The four-loop equal-mass banana integral}
As a last example, we discuss the evaluation of the master integrals for the four-loop equal-mass banana family, which is associated to a one-parameter family of CY threefolds.
A system of differential equations in canonical form for the equal-mass four-loop banana family was obtained in ref.~\cite{Pogel:2022ken}. The letters involve the periods of the underlying family of CY threefolds, as well as a second transcendental function related to the CY geometry, the so-called \emph{Yukawa coupling}, which can itself be expressed in terms of periods and quasi-periods. In addition, some of the letters are again expressed in terms of iterated integrals involving the periods and the Yukawa coupling.

\begin{figure}[t]
    \centering
    \includegraphics[width=0.49\textwidth]{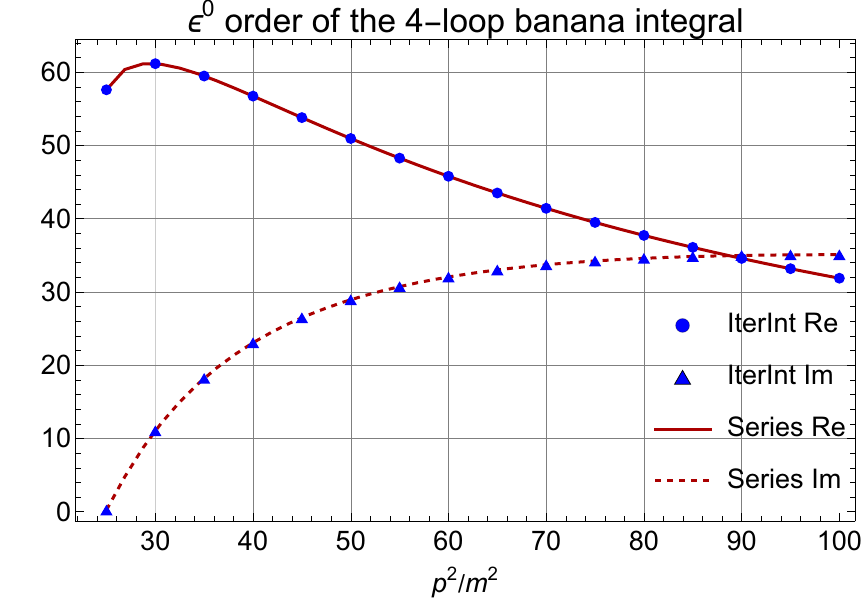}
    \includegraphics[width=0.49\textwidth]{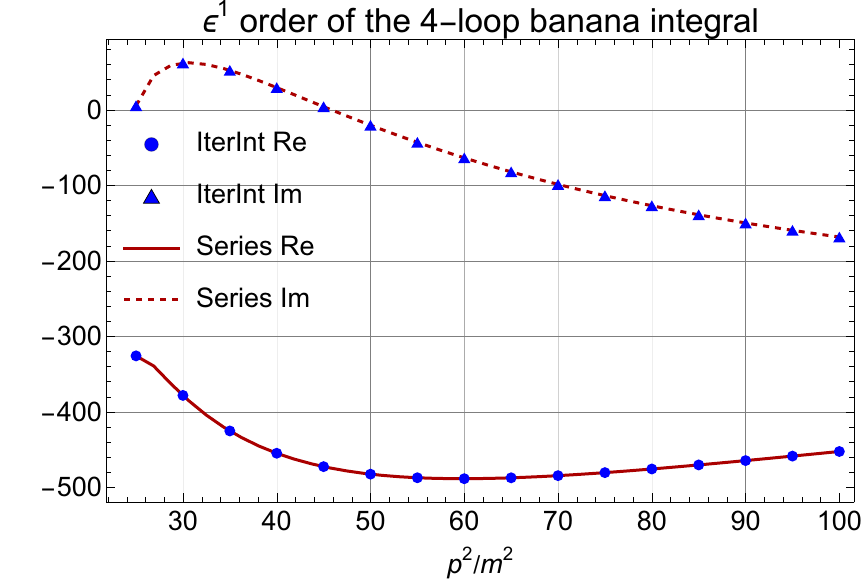}
    \caption{Numerical evaluation using \IterInt\ of the master integral $I_{11111}$ of the equal-mass 4-loop equal-mass banana family (blue markers). The red lines correspond to an independent evaluation of the integral. Note that the range on the $x$-axis corresponds to the region above threshold. Both the real (solid lines/circles) and imaginary (dashed line/triangles) parts are shown.}\label{fig:4loopComp}
\end{figure}

Unlike for the three-loop case, the periods of this family cannot be expressed in terms of periods of a family of elliptic curves.
From a mathematical perspective, the letters are much more complicated than for the previous examples, because they now involve various functions related to the CY geometry that cannot be reduced to functions attached to elliptic curves. From the perspective of numerical evaluation of the iterated integral with \IterInt, the complexity of the problem is not considerably higher than at two and three loops. Indeed, all the relevant letters admit converging series representations (cf., e.g., refs.~\cite{Bonisch:2020qmm,Klemm:2019dbm}),\footnote{The series representations typically have a finite radius of convergence, and one may need to analytically continue the periods to cover the whole kinematic space.} and so we can easily implement them into \IterInt\ and evaluate the iterated integrals. The results for the master integral $I_{11111}$ are shown in figure~\ref{fig:4loopComp}, and we again observe very good agreement with the numerics obtained from a truncated series representation of the result.

% !TEX root = main.tex
\section{Conclusion}\label{sec:conc}

In this paper, we have introduced a new method for the numerical evaluation of iterated integrals over general kernels. A main feature is that the user only needs to provide routines to numerically evaluate the integration kernels. Our code then immediately enables him/her to obtain numerical results for iterated integrals over these kernels by turning the iterated integrals into a first-order linear system which can be solved efficiently and with high precision using well established numerical libraries. The code is designed such that it can handle integrals that need to be shuffle-regulated, and it optimises the solution of the differential equations such as not to duplicate the evaluation of integrals.

We have implemented our algorithm into the public package \IterInt, both in \mathematica\ and in \cpp, and we have described the usage of the code in this paper. We note that this is not the first time that a public package is presented that allows one to evaluate iterated integrals that arise from differential equations for Feynman integrals, cf.,~e.g.,~ref.~\cite{Hidding:2020ytt}. \IterInt, however, uses a different strategy to evaluate the iterated integrals, and it is the first package to implement the combinatorial version of the shuffle regularisation from ref.~\cite{Brown:2014pnb}. As an illustration, and also as a means to validate our code and gauge its performance, we have compared the results of \IterInt\ to those obtained by \ginac\ for MPLs and eMPLs, and also for the first few orders in the dimensional regulator $\varepsilon$ of the banana integral with unit propagator powers. We foresee that \IterInt\ will have applications to solve also other classes of iterated integrals that arise from canonical differential equations satisfied by Feynman integrals.
\section*{Acknowledgements}
This work is funded by Deutsche Forschungsgemeinschaft (DFG, German Research Foundation) under the Research Unit FOR 5582 ``Modern Foundations of Scattering Amplitudes'' and under Germany’s
Excellence Strategy -- Cluster of Excellence ``Color meets Flavor'', EXC 3107 -- Project-ID 533766364.

\appendix

% !TEX root = main.tex

\section{Regularisation of eMPLs}\label{app:regularisation}

In section~\ref{sec:empls} we have already mentioned that there are two natural ways to regulate eMPLs via shuffle-regularisation. The first one consists in applying the prescription from section~\ref{sec:regularisation} directly to the eMPLs defined as iterated integrals in $z$. The second one is based on refs.~\cite{Broedel:2014vla,Broedel:2017jdo} and consists in introducing a tangential base-point at $w=1$, where $w$ is defined by $w=e^{2\pi i z}$. In this appendix we provide a means to translate between these two regularisation schemes. 

First, we note that we can always use the shuffle algebra properties to write any given eMPL in a form where the only function requiring regularisation is $\tilde{\Gamma}\qty({}^1_0; z, \tau)$. For concreteness, we denote in the following the regularised values of $\tilde{\Gamma}\qty({}^1_0; z, \tau)$ obtained by introducing a tangential base-point in the variables $z$ and $w$ by $\tilde{\Gamma}_z\qty({}^1_0; z, \tau)$ and $\tilde{\Gamma}_w\qty({}^1_0; z, \tau)$, respectively. Our goal is to show that these two schemes are simply related by a different choice of the parameter $v$. Note that this can also be seen as the result of applying the change of variables $w=e^{2\pi i z}$ to the respective tangent vectors.

We start from the integral representation of $\tilde{\Gamma}_z\qty({}^1_0; z, \tau)$, and we explicitly show the dependence on the parameter $v$:
    \begin{equation}
        \begin{aligned}
            \tilde{\Gamma}_z\qty({}^1_0; z, \tau) & =\int_0^z\dd{z'} \qty[g^{(1)}(z',\tau)-\inv{z'}]+\log\frac{z}{v}                                                                                                                                                                 \\
                           & =\int_0^z\dd{z'} \qty[g^{(1)}(z',\tau)-\frac{2\pi i}{e^{2\pi i z'}-1}+\frac{2\pi i}{e^{2\pi i z'}-1}-\inv{z'}]+\log\frac{z}{v}\,.
        \end{aligned}
    \end{equation}
    We now assume that Re$(z)\in (0,1)$. In that case, we have
    \beq
    \bsp
    \int_0^z\dd{z'} \qty[\frac{2\pi i}{e^{2\pi i z'}-1}-\inv{z'}] &\,= \log(e^{2\pi i z}-1) -\log(e^{2\pi i z})-\frac{i\pi}{2}-\log(2\pi z)\\
    &\,= \log(1-e^{2\pi i z}) -2\pi i z+\frac{i\pi}{2}-\log(2\pi z)\,.
    \esp\eeq
Comparing to eq.~\eqref{eq:eMPL_w_reg}, we see that we have, with $\log(-i) =-\tfrac{i\pi}{2}$:
\beq\bsp
\tilde{\Gamma}_z\qty({}^1_0; z, \tau) &\,=    \tilde{\Gamma}_w\qty({}^1_0; z, \tau) -\log(-2\pi iz) + \log\frac{z}{v}\,.
 \esp\eeq
 Hence, we see that $\tilde{\Gamma}_z\qty({}^1_0; z, \tau) =    \tilde{\Gamma}_w\qty({}^1_0; z, \tau) $, provided that we pick $v=-\tfrac{1}{2\pi i}$.

\bibliographystyle{JHEP}
\bibliography{lit.bib}

\end{document}